\documentclass[VANCOUVER,STIX1COL]{WileyNJD-v2}

\usepackage{graphicx}
\usepackage{textcomp}
\usepackage{xcolor}

\usepackage{subcaption}
\usepackage{enumitem}
\usepackage{ifthen}
\usepackage{soul}

\usepackage{hyperref}[hidelinks]

\usepackage{listings}
\usepackage{xspace}
\usepackage{booktabs}
\usepackage{multirow}
\usepackage{lscape} 
\usepackage{tcolorbox}
\usepackage{afterpage}

\def\BibTeX{{\rm B\kern-.05em{\sc i\kern-.025em b}\kern-.08em
    T\kern-.1667em\lower.7ex\hbox{E}\kern-.125emX}}

\newtheorem{termdefinition}{Definition}%
\newcommand{\tool}{\textsc{uncertainty-wizard}\xspace} %
\newcommand{\tfkeras}{\textit{tf.keras}\xspace} %

\newcommand{\changed}[1]{{\color{blue}#1}}

\articletype{Article Type}%

\received{18 October 2021}
\revised{03 October 2022}
\accepted{05 December 2022}

\begin{document}

\title{Uncertainty Quantification for Deep Neural Networks:\\ An Empirical Comparison and Usage Guidelines}

\author[1]{Michael Weiss*}

\author[1]{Paolo Tonella}

\authormark{Michael Weiss and Paolo Tonella}

\address[1]{
\orgdiv{Software Institute}, 
\orgname{Università della Svizzera italiana}, \orgaddress{\state{Lugano}, \country{Switzerland}}}

\corres{*\email{michael.weiss@usi.ch}}

\presentaddress{Via la Santa 1, 6962 Viganello, Switzerland}

\begin{abstract}

\end{abstract}

\abstract[Summary]{
Deep Neural Networks (DNN) are increasingly used as components of larger software systems that need to process complex data, such as images, written texts, audio/video signals. DNN predictions cannot be assumed to be always correct for several reasons, among which the huge input space that is dealt with, the  ambiguity of some inputs data, as well as the intrinsic properties of learning algorithms, which can provide only statistical warranties. Hence, developers have to cope with some residual error probability. An architectural pattern commonly adopted to manage failure prone components is the \textit{supervisor}, an additional component that can estimate the reliability of the predictions made by untrusted (e.g., DNN) components and can activate an automated healing procedure when these are likely to fail, ensuring that the Deep Learning based System (DLS) does not cause damages, despite its main functionality being suspended.

In this paper, we consider DLS that implement a supervisor by means of uncertainty estimation. After overviewing the main approaches to uncertainty estimation and discussing their pros and cons, we motivate the need for a specific empirical assessment method that can deal with the experimental setting in which supervisors are used, where accuracy of the DNN matters only as long as the supervisor lets the DLS continue to operate. Then we present a large empirical study conducted to compare the alternative approaches to uncertainty estimation. We distilled a set of guidelines for developers that are useful to incorporate a supervisor based on uncertainty monitoring into a DLS.}

\keywords{fault tolerance, software reliability, software testing, art neural networks, uncertainty quantification}

\jnlcitation{\cname{%
\author{M. Weiss} and
\author{P. Tonella}} (\cyear{2022}), 
\ctitle{Usage Guidelines for Deep Neural Network Uncertainty Quantification}, 
\cjournal{Software: Testing Verification and Reliability},
\cvol{2022}.}

\maketitle

\section{Introduction}
\label{sec:introduction}

Processing of large amounts of complex and unstructured data (e.g., images) to assign them automatically a label or to predict a value out of them has become a reality thanks to deep neural networks (DNNs). Moreover, the fast processors available even in small devices (e.g., mobile phones) allows deployment of DNNs even onto consumer hardware. The training process was boosted by the availability of fast and cheap GPUs. As a consequence, DNNs have spread across applications, also thanks to programming frameworks like \tfkeras (see \href{https://www.tensorflow.org/}{tensorflow.org}) that provide a user-friendly interface for DNN development to software engineers with possibly limited or no experience about DNNs.
Hence, we can find DNNs used in many \emph{Deep Learning based Systems (DLS)}, including self-driving cars, where they take as input sensor (e.g., camera) data and the output the controls for the car's actuators; health care, supporting doctors'  diagnosis and decision making; several web services for image transformation and analysis.

Regarded as software components, DNNs stipulate a probabilistic contract with the surrounding software system, and even if the error probability can be remarkably low, it cannot be assumed to be zero. Hence, DLS cannot ignore the possibility of errors, especially in safety-critical domains, and should instead account for the \emph{uncertainty} of the DNN about its own prediction. 

There are two main reasons for the uncertainty that affect DNNs~\cite{Kendall2017a}: (1) the entropy of the input; (2) inadequate training.
The first type of uncertainty is problem specific and cannot be avoided by definition (e.g., images taken in a given domain could be intrinsically noisy). The second can be reduced but in practice, it cannot also be eliminated, as the input space in interesting domains where DNNs are used is typically huge and involves a multitude of different conditions (e.g., weather or light conditions in which images are taken). It is not realistic to assume that training data have been collected so as to represent exhaustively all such conditions.

Given an input subject to high prediction uncertainty, the output that a DLS obtains from a DNN is just the predicted value. If the DLS blindly relies on such a prediction, it may fail if the prediction turns out to be incorrect. 
Notably, even well-tested, widely distributed DNNs are not prone to such failures: 
For example, a recent crash of a self-driving car was caused by the car not recognizing an overturned truck~\cite{Templeton2020Forbes}. Even in less safety-critical domains, such as image classification services available on the web, a prediction error can have serious consequences. A notable example that appeared in the news is the classification of the picture of a black person as a gorilla~\cite{Vincent2018GoogleFotos}. 
In summary, preventing DNN errors is quite important for society and will be even more in the future, but such a task is definitely challenging.

A  solution to this problem is the supervisor architectural design pattern: the DLS includes a \emph{supervisor} that monitors the DNN uncertainty for any given input at runtime, such that the system can ignore predictions for high-uncertainty inputs and can run a safe fallback process instead, such as stopping the self-driving car at the side of the street or delegating the classification of an image to a human~\cite{Stocco2020}.

The supervisor solution requires that the DNN making the prediction is also capable of assessing its own confidence (or uncertainty, the complement of confidence) in the prediction.
Indeed, there are  \emph{uncertainty-aware} types of DNNs, which support the deployment of such a supervisor. This paper aims at closing the gap between the alternative architectures that support uncertainty estimation and the practical usage of such estimators within an effective supervisor in a DLS.

An earlier version of this paper has been presented at the \emph{EEE International Conference on Software Testing, Verification and Validation 2021 (ICST 2021)}\cite{Weiss2021-EmpUnc}. 
This paper extends the previous version~\cite{Weiss2021-EmpUnc} both in its theoretical and empirical parts. 
In particular:

\begin{description}[noitemsep]
\item [Bayesian Neural Networks] We now  consider also Flipout-based Bayesian Neural Networks as an additional approach to uncertainty quantification.
\item [Optimal dropout rates] We investigate empirically optimal dropout rates in dropout-based uncertainty quantification. 
\item [Extended experimentation] We re-ran all experiments presented in the conference version of this paper and report new results for Flipout-based Bayesian DNNs.
\item [Expanded discussion] We have expanded the section with lessons learned and actionable guidelines.
\end{description}

The interested reader can find a description of \tool, a tool we presented alongside the conference version of this paper and which we used to collect all our results, in a dedicated testing tool paper published at ICST 2021~\cite{Weiss2021a}.

The paper is organized as follows: Section~\ref{sec:background} provides some background on the sources of uncertainty and ways to quantify it; Section~\ref{sec:approaches} describes the most widely adopted neural network architectures for uncertainty estimation; 
Section~\ref{sec:assessment} defines an empirical methodology for the experimental evaluation of supervisors; Section~\ref{sec:case_studies} describes the experiments conducted on four case studies, and consolidates them into lessons learned and actionable guidelines; Section~\ref{sec:related} discusses the related work, followed by conclusion in Section~\ref{sec:conclusion}.

\section{Background}
\label{sec:background}

In this section, we discuss the different root causes of DNN faults which can be understood as \emph{types of uncertainty}
and define the task of DNN fault prediction as a problem of uncertainty quantification.

\subsection{Sources of Uncertainty}

We distinguish between two types of uncertainty; uncertainty caused by a sub-optimal DNN model and uncertainty caused by randomness in the prediction target. 
A detailed discussion of these types is provided by Kendall \etal \cite{Kendall2017a} and by H\"ullermeier \etal \cite{hullermeier2021aleatoric}.

\begin{termdefinition}[Epistemic Uncertainty]
    Epistemic uncertainty denotes any uncertainty caused by suboptimal training or configuration of the model.
\end{termdefinition}

Epistemic uncertainty is sometimes also referred to as \emph{model uncertainty}.
There are many possible reasons for epistemic uncertainty, such as 
insufficient training data, which does not  represent the entire possible input space,
sub-optimal training hyper-parameters, and inadequate DNN architecture. 
In theory, epistemic uncertainty could be avoided, 
by providing good enough training data and optimal model configuration.
However, finding such optimal training configurations and data is impossible in most real-world applications,
as real input spaces, as well as the space of the possible hyper-parameters and architectural choices, are typically too large.

The second type of uncertainty, which not even an optimal training set and model configuration can avoid, is called \emph{aleatoric uncertainty}:

\begin{termdefinition}[Aleatoric Uncertainty]
    Aleatoric uncertainty is the uncertainty present in the true (unknown) distribution we are making predictions about.
\end{termdefinition}

Thus, aleatoric uncertainty can be seen as randomness, ambiguity, or entropy in the prediction target.
When predicting a random event, even an optimal model will make wrong predictions. 
As aleatoric uncertainty is independent of the model, but instead depends on the predicted data, it is also referred to as \emph{data uncertainty} or \emph{irreducible uncertainty}.
The only way to reduce aleatoric uncertainty in practice is thus through better inputs, containing more information useful for prediction (for example by using a larger input space).
Aleatoric Uncertainty can  be further distinguished between \emph{homoscedastic uncertainty}, 
where the uncertainty applies to all data, and \emph{heteorscedastic uncertainty}, 
where the uncertainty is more prevalent amongst some subsets of the data.

\def \unctypesubfigurewidth {0.18\linewidth}
\def \unctypeimgwidth {.8\linewidth}

\begin{figure}
\centering
\begin{subfigure}{\unctypesubfigurewidth}
  \centering
  \includegraphics[width=\unctypeimgwidth]{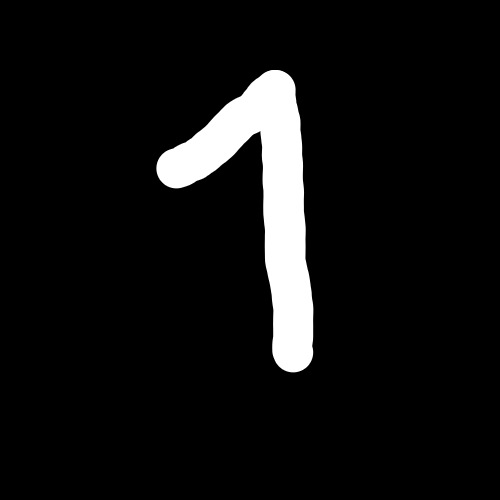}
  \caption{}
  \label{fig:u_type_nominal_1}
\end{subfigure}%
\begin{subfigure}{\unctypesubfigurewidth}
  \centering
  \includegraphics[width=\unctypeimgwidth]{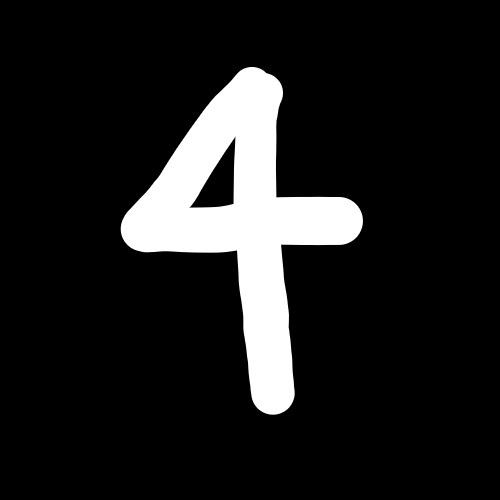}
  \caption{}
  \label{fig:u_type_nominal_4}
\end{subfigure}
\begin{subfigure}{\unctypesubfigurewidth}
  \centering
  \includegraphics[width=\unctypeimgwidth]{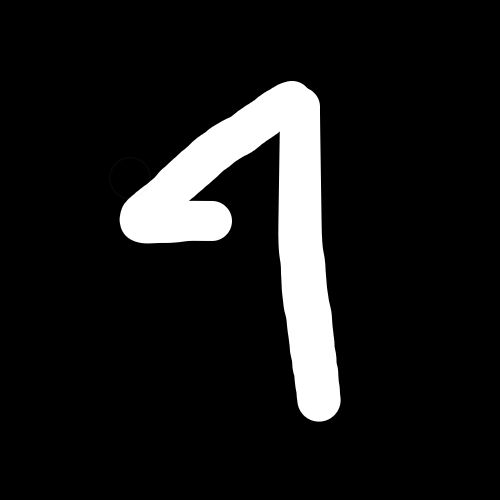}
  \caption{}
  \label{fig:u_type_aleatoric}
\end{subfigure}%
\begin{subfigure}{\unctypesubfigurewidth}
  \centering
  \includegraphics[width=\unctypeimgwidth]{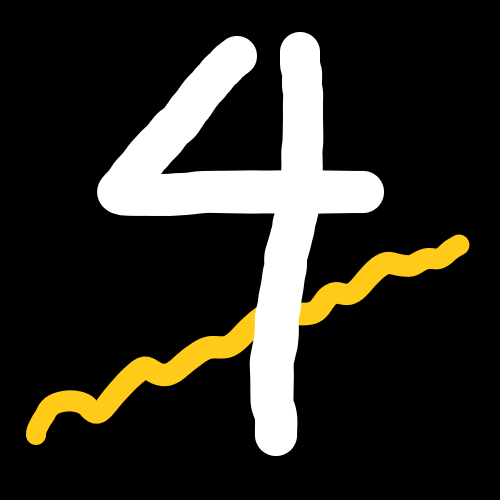}
  \caption{}
  \label{fig:u_type_epistemic_corrupted}
\end{subfigure}
\begin{subfigure}{\unctypesubfigurewidth}
  \centering
  \includegraphics[width=\unctypeimgwidth]{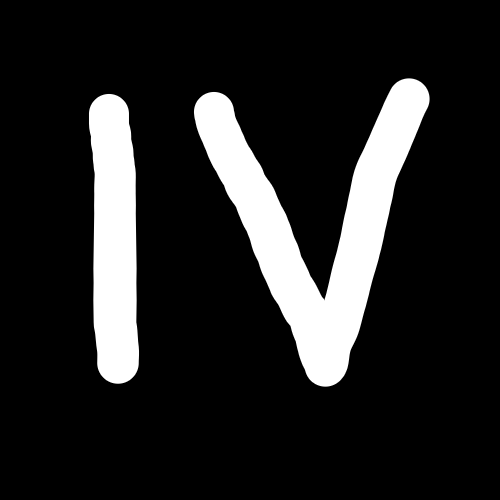}
  \caption{}
  \label{fig:u_type_epistemic_roman}
\end{subfigure}

\caption{
Examples of uncertainties in digits classification: 
(\subref{fig:u_type_nominal_1}) and (\subref{fig:u_type_nominal_4}) cause no uncertainty,
(\subref{fig:u_type_aleatoric}) causes aleatoric uncertainty, 
(\subref{fig:u_type_epistemic_corrupted}) and (\subref{fig:u_type_epistemic_roman}) cause epistemic uncertainty.
}
\label{fig:uncertainty_types}
\end{figure}
Figure \ref{fig:uncertainty_types} provides a visual example of the difference between aleatoric and epistemic uncertainty.
The input to a classifier DNN is the image of a handwritten digit to be recognized.
Figures \ref{fig:u_type_nominal_1} and \ref{fig:u_type_nominal_4} illustrate regular inputs with low uncertainty.
Figure \ref{fig:u_type_aleatoric} is an example of a figure with high heteroscedastic aleatoric uncertainty: 
Clearly, the image shows either a 1 or a 4, but it is impossible to say with certainty which one the writer intended.
Figure \ref{fig:u_type_epistemic_corrupted} illustrates a common reason for epistemic uncertainty:
A small perturbation of the image background, if not present in the training data, may lead the model to 
be incapable of predicting the correct label. 
Similarly, Figure \ref{fig:u_type_epistemic_roman} shows an unexpected input leading to epistemic uncertainty.
While the true label is unambiguously 4, a model which was not trained on roman number representations will not be capable 
to make a correct prediction.

\subsection{DNN Supervision}

\begin{figure}
    \centering
    \includegraphics[width=.8\linewidth]{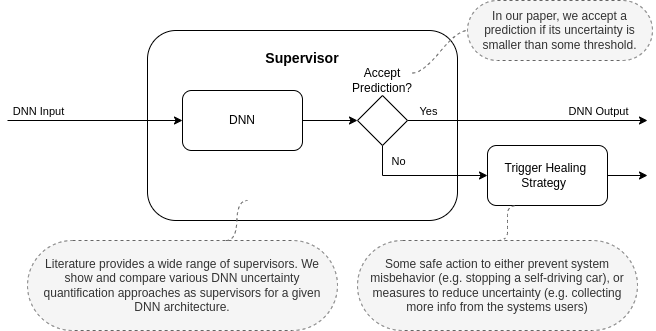}
    \caption{Supervised DNN}
    \label{fig:supervision}
\end{figure}

In practice, DLS are not exclusively faced with nominal inputs with clear ground truth, but also with invalid, out-of-distribution, ambiguous and adversarial inputs. 
This unavoidably leads not only to non-negligible uncertainty in the predictions but also to the presence of faulty predictions.
Fail-Safe DLS have to take countermeasures to make sure wrong predictions do not lead to critical system misbehavior.
To that extent, DNN supervision\cite{Henriksson2019, Henriksson2019a} (sometimes also called \emph{monitoring}~\cite{hendrycks2021unsolved}, \emph{prediction validation}~\cite{catak2022uncertainty}, \emph{input validation}~\cite{Wang2020} or \emph{misbehavior prediction}~\cite{Stocco2020} is one possible architecture to be used.
A schematic visualization of DNN supervision is provided in \autoref{fig:supervision}:
A DLS employs a \emph{supervisor} to detect inputs for which the output is likely to be wrong, and makes sure that such predictions are not used by subsequent components in the DLS.
Instead, the system may trigger a \emph{healing strategy} aiming to keep the DLS in a failure-free, safe state. 
In the case of a self-driving car, a healing strategy could be to immediately turn on warning lights and to slowly and safely stop the car. 
In cases in which a safe reaction is less time-critical, a system may also refuse to make any automated decisions but instead hand over the control to a human domain expert (e.g. in cases of AI-assisted diagnosis based on medical images, under high uncertainty, a system may not directly provide an assessment, but instead require  judgment from a medical doctor).
As a further example, faced with an ambiguous or unclear input leading to high uncertainty, a customer-facing DLS may ask the user to provide more information as part of their healing process, such that the next prediction attempt can be done with lower uncertainty.

\subsubsection{Choice of Supervisor}
Detecting inputs for which a DNN makes faulty predictions is still seen as one of the remaining open problems in machine learning safety~\cite{hendrycks2021unsolved}.
To that extent, recent literature provides a range of techniques that can be used for DNN supervision:
For example, various \emph{surprise adequacy} metrics~\cite{Kim2018, Kim2020, Kim2020a, Kim2021} can be used to detect out-of-distribution inputs based on the DNN inner activation patterns, allowing amongst other use cases  to predict DNN misclassifications~\cite{Weiss2021-SA}. 
Following the same motivation, but in a black-box fashion, variational autoencoders can similarly be used to detect inputs different from the training data~\cite{Stocco2020,stocco2020towardsContinuously}. 
A fundamentally different approach is taken in NIRVANA~\cite{catak2022uncertainty}, which combines the outputs of multiple uncertainty quantifiers in a support-vector machine~\cite{cortes1995svm} to predict when a DNN prediction is wrong.
DISSECTOR~\cite{Wang2020} combines per-layer classifications and predicts a misclassification dependent on the disagreement between per-layer classifications.
PURE~\cite{catak2021prediction} proposes a supervision framework for regression problems, specifically targeting self-driving cars.
The interested reader can refer to the existing literature for comparisons between these, and more, types of DNN supervisors~\cite{berend2020cats, weiss2022forgotten, Ferreira2021BenchmarkingSafetyMonitors}.

However, the most obvious implementation of a supervisor is to use the DNN's uncertainty: If, and only if, the DNNs uncertainty is lower than some threshold a prediction can be trusted and accepted for future processing. 
To that extent, existing literature proposes a range of uncertainty quantification techniques (explained below). 
This paper considers the use of uncertainty as a supervisor and compares the performance of various uncertainty quantification techniques  in such DNN supervision setting. 
Using uncertainty as supervisor has clear advantages over other supervisors: Amongst others, uncertainty quantification techniques are intensively studied and widely used (see e.g. survey by Jospin et. al. \cite{Jospin2020}), typically do not require the implementation of large additional system components, or -- where needed -- are implemented in a range of uncertainty quantification frameworks (survey provided by Pintz et. al.\cite{pintz2022uncertaintyToolsSurvey}), including uncertainty-wizard~\cite{weiss2021uwiz}, which allows uncertainty quantification in an almost transparent way.

It is worth noticing that, with few notable exceptions~\cite{Stocco2020, stocco2020towardsContinuously}, supervisors are evaluated at the model-level, 
despite their ultimate goal being to prevent system-level failures. 
While in some domains (such as image labeling) the two levels (model vs system) are identical, in many applications there is a  correlation, but not an equivalence between them.
Of course,  bugs in the code and hardware problems can also lead to a system failure, even when the DNN does not make any mispredictions.
We refer to the related literature for an in-depth discussion of the difference between model- and system-level testing in DLS~\cite{Riccio2020, codevilla2018offline, haq2020comparing, haq2021can}.
However, it is important to remark the role of DNN supervisors when considering model- vs system-level failures: use of a DNN supervisor, which operates at the model-level, increases the overall system reliability, circumventing the problem that in most cases model-level faults cannot be prevented at runtime. In other words, thanks to DNN supervisors, model-level faults, whose probability cannot be realistically reduced to zero in production, do not propagate at the system level as system failures if supervisors are adopted to monitor and wrap the execution of the DNN model. On the other hand, a complex system that includes DNN, as well as traditional software components, will need a properly designed self-healing policy to ensure that complex interactions between possibly faulty components do not compromise the overall system safety~\cite{AbdessalemPNBS18}, even in the presence of supervisors.

\subsection{Uncertainty Quantification}
Ideally, instead of predicting a single value as output, a DNN should calculate a probability density function for regression problems or a likelihood for every outcome in a classification problem.
As such, every outcome would have its uncertainty quantified (e.g., by the variance of the output probability distribution).
We will discuss models capable of calculating such outputs in Section \ref{sec:approaches}.
However, for the scope of this paper, we consider a less general formulation of uncertainty quantification,
which is sufficient for network supervision~\cite{Riccio2020}:

\begin{termdefinition}[Uncertainty Quantification]
    Uncertainty Quantification (UQ) is the task of calculating a scalar metric
    strictly monotonically increasing in the likelihood (for classification tasks) or severity (for regression tasks)
    of a deviation between the DNN prediction and the ground truth, given a particular input and the DNN used to do the prediction.
\end{termdefinition}

We are thus limiting our interest to the correctness of the chosen prediction,
as opposed to the distribution of all possible predictions in the output space.
The supervisor will reject inputs for which the uncertainty is above a certain threshold.

Consistently, we define \emph{confidence} as the opposite of uncertainty, 
s.t. a confidence metric is supposed to strictly monotonically decrease in the likelihood or severity of a prediction error.

\section{Uncertainty-Aware Neural Networks}
\label{sec:approaches}
\def \spacebetweenapproachrows {0.1cm}

\begin{table*}[t]
\centering
\resizebox{\textwidth}{!}{
\begin{tabular}{@{}lllllllll@{}}
\toprule
\textbf{\begin{tabular}[c]{@{}l@{}}Uncertainty-\\ aware DNN\end{tabular}}            & \textbf{\begin{tabular}[c]{@{}l@{}}Classifi-\\ cation\end{tabular}} & \textbf{\begin{tabular}[c]{@{}l@{}}Regre-\\ ssion\end{tabular}}& 
\textbf{\begin{tabular}[c]{@{}l@{}}BNN\end{tabular}} & \textbf{\begin{tabular}[c]{@{}l@{}}Custom\\ Layers\end{tabular}} & \textbf{\begin{tabular}[c]{@{}l@{}}Training\\ Effort\end{tabular}} & \textbf{\begin{tabular}[c]{@{}l@{}}Prediction\\ Effort\end{tabular}} & \textbf{\begin{tabular}[c]{@{}l@{}}Main \\ Advantage\end{tabular}}                                     & \textbf{\begin{tabular}[c]{@{}l@{}}Main \\ Disadvantage\end{tabular}}                                              \\ \midrule

\vspace{\spacebetweenapproachrows}
\textbf{\begin{tabular}[c]{@{}l@{}}Pure BNN\end{tabular}}                                                                                                                                   & Yes                                                             & Yes                                                                          & Yes          & \begin{tabular}[c]{@{}l@{}}Weight \\ distributions\end{tabular}  & Intractable     & \begin{tabular}[c]{@{}l@{}}High or \\ intractable\end{tabular}  & \begin{tabular}[c]{@{}l@{}} (Hypothetical) \\ most general BNN \end{tabular}                                        & \begin{tabular}[c]{@{}l@{}}Intractable, \\ hence not practical\end{tabular}                                          \\
\vspace{\spacebetweenapproachrows}
\textbf{\begin{tabular}[c]{@{}l@{}}(Impure) \\PW-BNN\end{tabular}}                                                                                                                                   & Yes                                                             & Yes                                                                          & Yes          & \begin{tabular}[c]{@{}l@{}}Weight \\  distributions\end{tabular}  & \begin{tabular}[c]{@{}l@{}}High*\end{tabular}      & \begin{tabular}[c]{@{}l@{}}High*\end{tabular}        & \begin{tabular}[c]{@{}l@{}}Theoretically \\ well founded\end{tabular}                                         & \begin{tabular}[c]{@{}l@{}}Custom layers, \\ difficult to train\end{tabular}                                          \\
\vspace{\spacebetweenapproachrows}
\textbf{MC Dropout}                                                                                                                                & Yes                                                             & Yes                                                                          & Yes          & Dropout                                                          & Minimal                                                                & High                                                                 & \begin{tabular}[c]{@{}l@{}}Fastest BNN\\ approximation\end{tabular}                                    & \begin{tabular}[c]{@{}l@{}}Sampling for \\ predictions is costly\end{tabular}                                      \\
\vspace{\spacebetweenapproachrows}
\textbf{\begin{tabular}[c]{@{}l@{}}Deep \\ Ensembles\end{tabular}}                                                                                 & Yes                                                             & Yes                                                                          & No*           & No                                                               & High                                                               & High                                                                 & \begin{tabular}[c]{@{}l@{}}Good practical results,\\ no major architecture\\ requirements\end{tabular} & \begin{tabular}[c]{@{}l@{}}Computationally and \\ storage intensive\end{tabular}                                              \\
\vspace{\spacebetweenapproachrows}
\textbf{\begin{tabular}[c]{@{}l@{}}PPNN-Softmax\\ based\end{tabular}} & Yes                                                                 & No                                                                                                                                 & No           & Softmax                                                          & Miminal                                                            & Minimal                                                              & Fast and Simple                                                                                        & \begin{tabular}[c]{@{}l@{}}Misleading and\\ no regression\end{tabular}                                                \\
 \bottomrule
\end{tabular}
}
\caption{Overview of popular uncertainty-aware DNN techniques (* = approach dependent) }
\end{table*}
\label{tab:supervision_approaches}

While regular DNNs, also called \emph{Point-Prediction DNNs (PPNN)}, predict a single scalar value for every output variable, from the perspective of uncertainty quantification it would be preferable if the DNN calculated a distribution over possible output values, which would allow the extraction of the network's confidence in the prediction. 
In this section we discuss the various DNN architectures that allow the calculation of such distributions:
first, we  provide an overview on \emph{Bayesian Neural Networks (BNN)}, which are the standard approach in the machine learning literature about uncertainty quantification, and ways to build them.\footnote{
For a more detailed discussion of BNN, we recommend the comprehensive and usage-oriented article by Jospin \etal ~\cite{Jospin2020}.
}
Second, we discuss other approaches, which typically compromise some theoretical disadvantages in exchange for some practical advantages when compared to BNNs, and which showed good performance.\cite{Ovadia2019}
An overview is provided in \autoref{tab:supervision_approaches}.

\subsection{Pure Bayesian Neural Networks}
BNNs are neural networks with probabilistic weights, instead of scalar weights as in PPNN, and are represented as probability density functions.
To train a BNN, first, a prior distribution $p(\theta)$ over weights $\theta$ has to be defined. 
Then, given some data $D$, the posterior distribution $p(\theta|D)$, i.e., the trained BNN is inferred using Bayes rule: 
\begin{equation}
p(\theta|D) = \frac{p(D|\theta) p(\theta)}{p(D)}  
= \frac{p(D|\theta) p(\theta)}{\int p(D|\theta) p(\theta) d\theta}
\end{equation}

Since weights are probabilistic, any  output of the BNN is also probabilistic and thus allows a statistically well-founded uncertainty quantification.
Besides that, these BNNs in their pure form  have other advantages over PPNN, among which: 
they are robust to overfitting~\cite{Neal1992} and allow to distinguish between epistemic and aleatoric uncertainty~\cite{Jospin2020}.

\subsection{(Impure) Probabilistic Weights Bayesian Neural Networks}
BNNs in the pure form presented above quickly become intractable due to the large number of integrations required.
There has been a long search for mechanisms to approximate the true posterior in BNN, with papers dating back to the early 1990s~\cite{MacKay1992, Neal1992}. 
In this paper, we summarize models where at least some of the weights are modelled as distributions. An example is \emph{Probabilistic Weights BNN (PW-BNN)}.
The development of such models has recently been facilitated by the release of software libraries with ready-to-use PW-BNN layers, such as \textsc{tensorflow-probability}\footnote{https://www.tensorflow.org/probability}, which offers, amonst others Variational Inference Layers,
Reparameterization Layers\cite{Kingma2014} and Flipout Layers\cite{Wen2018}.

While models built with such layers are not \emph{pure} BNN, i.e., they only approximate the true posterior distribution, the nondeterministic nature of their weights still  captures well the  essence of BNN.
Arguably, the main practical disadvantages of non-pure PW-BNN are thus not their impurity, but the problems which arise from having probabilistic weights. 
For example, as for any bayesian inference, a prior distribution has to be chosen based on the developers' insight of the modeled problem.
In deep learning, where model weights have little if any practical meaning, the choice of a prior distribution is difficult. 
In addition, despite many advances in the field, not only do some consider even recent approaches to not scale sufficiently well~\cite{Ilg2018},
but more scalable approaches have also been shown to  outperform PW-BNN on practical benchmarks~\cite{Ovadia2019}.
Hence, it is not surprising that in recent work PW-BNNs are still not  widely used in practice~\cite{Jospin2020}.

\subsection{MC-Dropout based Bayesian Neural Networks}
In their very influential paper\footnote{More than 2300 citations in the four years since its publication, according to \href{https://scholar.google.com/}{Google Scholar}.}, Gal \etal~\cite{Gal2016} proposed to approximate BNNs through regular PPNNs.
Many PPNNs use \emph{dropout layers} for regularization. During training, in a dropout layer, each neuron activation can be set to 0 with some probability $p$. This helps to avoid overfitting and can lead to better model performance in general~\cite{Srivastava2014}.
In their paper, Gal \etal have shown that a PPNN trained with dropout enabled can be interpreted as a BNN,
due to the variation induced by the randomized dropout layers.
While traditionally the dropout functionality is disabled at prediction time, a dropout-based BNN keeps the random dropping of neuron activations
enabled even during predictions and calculates an output distribution by Monte-Carlo sampling of the results in multiple randomized  predictions.
This approach is thus often referred to as \emph{MC-Dropout}.

Despite its high popularity, MC-Dropout  is not without criticisms: 
Osband \cite{Osband2016} claimed the inability of MC-Dropout to capture every type of uncertainty and
several papers have shown that it can be outperformed by other approaches~\cite{Lakshminarayanan2017, Ovadia2019}.
Also, despite the fact that \textit{"the forward passes can be done concurrently, resulting in constant running time"} \cite{Gal2016},
concurrent processing may not be possible in resource-constrained environments, making MC-Dropout clearly slower than a single prediction of the corresponding PPNN.

\subsection{Deep Ensemble Neural Networks}
Lakshminarayanan \etal proposed a fundamentally different approach to quantify uncertainty called \emph{Deep Ensembles}, or \emph{Ensemble} for short, i.e., a collection of multiple \emph{atomic models} for the same problem,
differing only in their  initialization and trained independently. 
For every input, a prediction would be made on every atomic model. 
The predictive distribution could then be inferred from these samples.
While deep ensembles are not inherently BNNs, it is possible to interpret them as BNN after applying some minor changes to the parameter regularization~\cite{Pearce2020}.
Nonetheless, even in their plain form, they have been shown to outperform MC-Dropout on the task of uncertainty quantification~\cite{Lakshminarayanan2017, Ovadia2019}.

Deep Ensembles are, compared to a single PPNN, slow to train if executed sequentially and memory intensive if used concurrently.
This may prevent the use of ensembles in some constrained environments.
A variety of improvements and modifications have been proposed for Deep Ensembles (Ilg \etal \cite{Ilg2018} provide a good overview of them).

\subsection{Point Predictor Classifiers}
DNNs used for classification typically have a \emph{softmax} output layer, including one output neuron per class. 
The output for each class is between $0$ and $1$, with the sum of all outputs being exactly $1$.
These outputs are often interpreted as probability distributions over the classes
and are used for network supervision by means of the following quantifiers:

\begin{termdefinition}[Max-Softmax (SM)] The highest softmax score is used as confidence quantification 
   (also referred to as \emph{Vanilla}~\cite{Ovadia2019} or \emph{Softmax Prediction Probability}~\cite{Berend2020} quantification).
\end{termdefinition}

\begin{termdefinition}[Prediction Confidence Score (PCS)~\cite{Zhang2020}]
    The difference between the two highest softmax outputs is used as confidence quantification.
\end{termdefinition}

\begin{termdefinition}[Softmax-Entropy (SME)]
    The entropy over the outputs of the softmax layer is used as  confidence quantification.
\end{termdefinition}

These quantifiers are often criticized as a poor approach to uncertainty quantification:
As opposed to BNNs and BNN approximations, PPNN-based quantifiers are not theoretically well-founded
and can be shown to severely overestimate a network's confidence~\cite{Gal2016}.
As a further disadvantage, such a PPNN-based approach can not be directly applied to regression problems.

\subsection{Inferring Prediction and Uncertainty from Samples}
In MC-Dropout and Deep Ensembles, samples need to be aggregated into a point prediction and uncertainty quantification,
and the literature provides a variety of quantifiers able to do so.
In this Section, we first discuss the quantifiers applicable to regression problems, and then the ones for classification problems.
We provide a simple-to-use implementation of all these quantifiers as part of \tool~\cite{weiss2021uwiz}.

\paragraph{Regression Problems}

In their proposition of MC-Dropout, Gal \etal~\cite{Gal2016} propose to use the average of the observed samples as prediction,
and their predictive variance as uncertainty:
\begin{termdefinition}[Predictive Variance]
    For what regards DNN supervision,  \emph{predictive variance} is defined as the sample variance over the observed samples.
\end{termdefinition}
An alternative approach was proposed by Lakshminarayanan \etal\cite{Lakshminarayanan2017}.
For their deep ensembles, they propose that the atomic models are adapted s.t. they have a second output variable for every regression output that predicts the variance~\cite{Nix1994}. 
The uncertainty of the ensemble can then be quantified by averaging these variances. 

\paragraph{Classification Problems}
The following quantifiers are proposed to derive an overall prediction and uncertainty:

\begin{termdefinition}[Mean-Softmax (MS)]
    The overall prediction is the class with the highest sum of softmax scores over all samples
    and the corresponding confidence is the average softmax score of this class over all samples.
\end{termdefinition}

MS has been proposed and is often used with Deep Ensembles. It is thus also called \emph{ensembling}\cite{Jospin2020}.
Three other quantifiers have been proposed to be used with MC-Dropout\cite{Gal2016a, Gal2016, Michelmore2018}. 
We refer to Gal, 2016~\cite{Gal2016}, Section 3.3.1 for a precise and formal description, and for examples of the following three uncertainty quantifiers:
\begin{termdefinition}[Variation Ratio (VR)]
    The percentage of samples for which the overall chosen class is \textbf{not} the class with the highest softmax output
\end{termdefinition}

\begin{termdefinition}[Predictive Entropy (PE)]
    The average amount of information in the predicted distributions.
\end{termdefinition}

\begin{termdefinition}[Mutual Information (MI)]
    The mutual information metric (from Information Theory) between a prediction and the model's posterior distribution.
\end{termdefinition}

\section{Supervised Neural Network Assessment}
\label{sec:assessment}

Network supervision can be viewed as a binary classification task: 
\emph{malicious samples}, i.e., inputs that lead to a misclassification (for classification problems) or to severe imprecision (in regression problems) are positive samples that have to be rejected.
Other samples, also called \emph{benign samples}, are negative samples in the binary classification task.
An uncertainty-based supervisor accepts an input $i$ as a benign sample if its uncertainty $u(i)$ is lower than some threshold $t$.
The choice of $t$ is a crucial setting, as a high $t$ will fail to reject many malicious samples (false negatives) and a low $t$ will cause too many false alerts (false positives).

Differently from standard binary classification tasks, the choice of $t$ for network supervision cannot rely on the optimization of an aggregate metric that accounts for both false positives and false negatives, such as the F1-metric, because negative samples are either completely unknown at training time, or, in case some negative samples are known, they cannot be assumed to be representative of all unknown execution conditions that will give raise to uncertainty at runtime.
Hence, the choice of $t$ is solely based on the false positives observed in the validation set.
In practice, given the uncertainties measured on the validation set, $t$ shall ensure that at runtime, under  conditions similar to those observed on  test data, only an acceptable \emph{false positive rate} $\epsilon$ is expected to occur~\cite{Stocco2020}.

Existing metrics allow the individual assessment of the supervisor's performance separately from the assessment of the model performance\cite{Henriksson2019}. However, such measurements do not take into account the interaction between the two: since the output of the model is not used by the DLS when the supervisor activates the fail-safe procedure ($u(i) \geq t$), it does not make any sense to evaluate the performance of the model in such scenarios. For this reason, we propose a new approach for the joint assessment of model and supervisor, which we call \textit{supervised metrics}. In the next two sections, we first summarize the state-of-the-art metrics for the separate, individual assessment of model and supervisor, followed by a description of our proposal for a new joint assessment approach.

\subsection{Existing Metrics for the Individual Assessment of Model and Supervisor}

There are well established metrics for the individual assessment of  performance of a model $m$. These are  based on some \emph{objective function} $obj(I,m)$, such as \emph{accuracy (ACC)} (for classifiers) or \emph{mean-squared error (MSE)} (for regression models), computed on a test dataset $I$.\footnote{$obj$ does not have to be the same function used to optimize the DNN during training.}

Classically, the supervisor's performance would be assessed individually using  performance metrics designed for binary classifiers. For a given $t$, the available metrics include \emph{true positive rate (TPR)}, \emph{false positive rate (FPR)}, \emph{true negative rate (TNR)} and \emph{false negative rate (FNR)}, \emph{$F_1$} score and \emph{accuracy (ACC)}. 
To use these metrics with regression problems, an \emph{acceptable imprecision} would have to be defined, allowing to divide inputs into benign (negative cases) and malicious (positive cases).
Alternatively, the effect of the predictions on the overall DLS system could be monitored to only treat inputs leading to system failures as malicious ones~\cite{Stocco2020}.

There are also existing, classical metrics to assess a binary classifier independently of the threshold $t$.
For instance, the \emph{average precision score (AVGPR)} computes
the average of the precision values obtained when varying $t$, weighted by the recall measured at $t$. 
Another popular threshold independent metric is the \emph{area under the receiver operating characteristic curve (AUROC)}\cite{Stocco2020, Berend2020, Michelmore2018}. 
When individually assessing the performance of a supervisor, AVGPR should be preferred over AUROC as, amongst other advantages \cite{Davis2006}, it is better suited for the unbalanced datasets~\cite{Saito2015} typically observed during malicious input detection.
Threshold independent analysis of the supervisor in a regression problem is  straightforward and can be done using point-biserial correlation between the quantified uncertainty and the observed prediction error, given some objective function (e.g. MSE). 

Independent analysis of model's performance, considered in isolation without supervision, and of the supervisor's performance, again in isolation, results in measurements that do not capture the overall goal of the interaction between supervisor and model under supervision: ensuring high model performance on the samples considered safe by the supervisor, while minimizing the amount of samples  regarded as unsafe. 
To capture such goal, we propose novel metrics for joint model and uncertainty quantification assessment in a supervised DLS.

\subsection{Supervised Metrics for the Joint Assessment of Model and Supervisor}
When considering model and supervisor jointly, we can still use the objective function adopted to assess the model in isolation, but  we evaluate it in a supervised context:
given a test set $I$ and an objective function $obj(I,m)$ for model $m$ with uncertainty quantifier $u$ and supervisor threshold $t$, the \textit{supervised objective function} $\overline{obj}(I,m)$ is defined as:
$$
\overline{obj}(m,I) = obj(\{i \mid i \in I \text{ and } u(i) < t \})
$$
i.e., the objective is applied only to the subset of inputs which is accepted by the supervisor.
By decreasing $t$, assuming  that the cardinality of the resulting subset of inputs remains big enough to calculate a statistically significant $\overline{obj}(I)$, we may generally get higher values of the supervised objective function $\overline{obj}(I)$.
However, such high values of $\overline{obj}(I)$ are likely associated with a high false alarm rate of the supervisor.
Thus, any $\overline{obj}(I)$ should always be regarded in conjunction with the acceptance rate $\Delta_u(I)$ of the supervisor:
$$
\Delta_u(I) = \frac{\mid \{i \mid i \in I \text{ and } u(i) < t \} \mid }{\mid I \mid}
$$

Similar to the popular $F1$ score, the following combination of these two metrics allows to capture the effectiveness of the collaboration between supervised model and supervisor:
\begin{termdefinition}[S-Score]
The $S_1$-Score measures the harmonic mean of a supervised objective function, normalized between zero and one, and the supervisors' acceptance rate, given a model $m$, uncertainty quantifier $u$ and test set $I$ as defined above, as 
$$
S_1(m,u,I) = \frac{2}{\frac{obj^+ - obj^-}{\overline{obj}(I,m) - obj^-} + \Delta_u(I)^{-1}}
$$
\end{termdefinition}
where $obj^-$ and $obj^+$ are the lower and upper bounds used for normalization of the objective function.
For classifiers, if  accuracy is the objective function,  $obj^- := 0$ and $obj^+ := 1$. For regression problems, or more generally for unbounded objective functions, $obj^-$ and $obj^+$ have to be estimated empirically (e.g., based on the empirical distribution of the objective function values), independently from $m$ and $u$.

The $S_1$ scores weights $\Delta_u(I)$ and $\overline{obj}(m,I)$ equally. 
Equivalent to the popular $F_1$ score, other $S_\beta$ scores can be used, where $\beta > 0$ is the weighting parameter~\cite{Rijsbergen1979}:

$$
S_\beta(m,u,I) = (1 + \beta^2)\cdot \frac{\frac{\overline{obj}(I,m) - obj^-}{obj^+ - obj^-} \cdot \Delta_u(I)}{(\beta^2 \cdot \frac{\overline{obj}(I,m) - obj^-}{obj^+ - obj^-}) + \Delta_u(I)}
$$

Summarized, $S_{\beta}(m, u, I)$ allows comparing the performance of various supervised models $m$ using various supervisors $u$ for a given test set $I$, where constant $\beta$ defines the priorities (i.e., risk-aversion) in the trade-off between high acceptance rate $\Delta_u(I)$ and high supervised performance  $\overline{obj}(I)$.

\section{Case Studies}

\label{sec:case_studies}
We assess the uncertainty quantification capabilities of point predictors, deep ensemble, PW-BNN and MC dropout using different quantifiers. 
The \textit{goal} of our empirical evaluation is twofold:
\begin{enumerate}
    \item To assess the usefulness of different techniques for uncertainty
    quantification when used as DNN supervisors.
    \item To provide clear and actionable guidelines to researchers and practitioners who aim to use uncertainty quantification for their DNNs.
\end{enumerate}

Our case studies build around four different types of uncertainty-aware DNNs per subject, 
all based on publicly available DNN architectures:
\begin{itemize}
    \item \textbf{Point-Predictors}, where we use the publicly available DNN architecture without applying any changes.
    \item \textbf{MC-Dropout}, where we use the same models trained as Point-Predictors, as all the considered models already had at least some dropout layers.
    \item \textbf{Deep Ensembles}, where we used the publicly available DNN architecture to create 50 atomic models per ensemble.
    \item \textbf{Flipout} as an example for a PW-BNN, where we replaced convolutions and dense layers with corresponding flipout layers~\cite{Wen2018}, and where we removed all dropout layers from the model architecture. 
\end{itemize}

\subsection{Research Questions}

We consider the following research questions:

\def \rqvspace {\vspace{10pt}}

\rqvspace
\noindent
\textbf{RQ\textsubscript{1} (effectiveness):}
\textit{How effective are  supervisors at increasing the supervised model's performance?}

This is the key research question of our empirical study, since the main hypothesis behind supervisors is that they can prevent usage of a model when its performance is predicted to be low. Hence, we expect an increase of the supervised model's performance $\overline{obj}$ as compared to the unsupervised one $obj$.

\rqvspace
\noindent
\textbf{RQ\textsubscript{2} (comparison):}
\textit{Is there a supervisor and quantifier type which yield optimal performance across  subjects and across alternative choices of the uncertainty threshold?}

We consider four types of uncertainty estimators, Point Predictors, MC-Dropout, Ensemble and Flipout, and several uncertainty quantifiers, respectively [SM, PCS, SME], [VR, PE, MI, MS], [VR, PE, MI, MS], [VR, PE, MI, MS] (see Section~\ref{sec:approaches}). We want to investigate whether any combination of estimator type and quantifier dominates all the others in terms of $S_1$-score. To investigate how performance changes with the uncertainty threshold $t$, we consider different acceptable rates $\epsilon$ of false positives on the nominal data and we compute the threshold $t$ that ensures such FPR on the validation set, so that we can compare alternative estimators/quantifiers at equal FPR on the validation set of the nominal data.

\rqvspace
\noindent
\textbf{RQ\textsubscript{3} (sample size):}
\textit{How many samples are required in stochastic and ensemble models to get reliable uncertainty quantification?}

Since the main cost associated with the usage of MC-Dropout and Ensemble is the collection of multiple samples for each individual prediction, we want to understand what is the minimum sample size that ensures good performance of each different supervisor. In particular, we study the convergence of  supervised accuracy to its asymptotic value as the sample size is increased.

\rqvspace
\noindent
\textbf{RQ\textsubscript{4} (sensitivity):}
\textit{How sensitive are supervisors to changes in their main hyperparameters?}

With this research question we want to understand whether the choice of hyperparameters is critical to get optimal performance, or  on the contrary if they can be chosen in the neighbourhood of the optimal choice with minor impact on the resulting performance of the supervisor.
For Point Predictors, we consider the number of training epochs as the main hyperparameter;
for MC-Dropout, the number of training epochs and the number of samples; for Ensemble, the number of training epochs and the number of atomic models. We measure the standard deviation of the supervised objective function (e.g., supervised accuracy) in the neighbourhood of each hyperparameter choice, so as to identify the regions where such standard deviation is low.

\rqvspace
\noindent
\textbf{RQ\textsubscript{5} (dropout rate):}
\textit{How does the choice of dropout rate influence MC-Dropout's uncertainty quantification capability?}

In MC-Dropout, the dropout rate is an important hyperparameter to steer the amount of randomness present during training and sampling.
Intuitively, one may consider to choose an artificially high dropout rate to make sure there is \emph{'enough'} randomness for UQ.
With RQ\textsubscript{5}, we want to understand whether practitioners should indeed dedicate special care when choosing a dropout rate with uncertainty quantification in mind.

\subsection{Subjects}

We use the following classification problems as case study subjects, aiming to increase diversity and practical relevance. 
\begin{description}
    \item [Mnist\cite{LeCun1998}] Classification of hand-written digits, formatted as small grayscale images. 
    This is the most popular dataset in machine learning testing~\cite{Riccio2020}, and a relatively easy problem, where even simple models achieve high accuracy. 
    We took the DNN architecture from a Keras tutorial~\cite{KerasMnistModel}.
    
    \item [Cifar10\cite{Krizhevsky2009}] Classification of colored images into ten different classes. It is also very popular in DLS testing~\cite{Riccio2020} and it represents a more challenging task than Mnist.
    We use the model architecture proposed in the Brownlees Cifar10 tutorial~\cite{BrownleeCifarModel}.
    
    \item [Traffic~\cite{Serna2018}] Classification of images of European traffic signs~\cite{Segvic2010, Bonaci2011, Mathias2013, Timofte2014, Belaroussi2010, Stallkamp2012, Grigorescu2003, Larsson2011}.
    The different sources the data was collected from, combined with the fact that the dataset is unbalanced and many images are of bad quality, reflect a quite realistic, high-uncertainty setup. Since traffic sign recognition is a core component of self-driving cars, this is also a very interesting case study from the software and system engineering point of view.
    The model architecture we use was proposed alongside the release of this dataset~\cite{Serna2018}.
    
    \item [Imagenet~\cite{Deng2009} (Pretrained)] Image classification problem with as many as 1,000 classes.
    We use  eight pre-trained \emph{Efficientnet} models~\cite{Tan2019}.
    As for this subject we rely on pre-trained models (which include dropout layers), we can test them only 
    as MC-Dropout and Point Predictor models, but not as Ensembles or Flipout models.
\end{description}

\subsection{Experimental Setup}

Except for the pre-trained ones, models were trained for 200 epochs. 
After every epoch, we assessed the models' performance on both a nominal and an out-of-distribution (OOD) dataset, for every quantifier.
To do so, we used Mnist-c~\cite{Mu2019} as OOD test set for Mnist and the color-image transformations
proposed by Hendrycks~\cite{Hendrycks2018} to generate OOD samples for the other subjects.
We used three different thresholds, calculated on the nominal validation set to ensure the lowest possible FPR above $\epsilon$, with $\epsilon\in\{.01,.5,.1\}$ respectively.
To measure the sensitivity to the number of samples, quantifiers of deep ensembles were assessed with every number of atomic models between 2 and 50,
Similarly, MC-Dropout and Flipout models were assessed on every number of samples between 2 and 100.
To answer RQ4 (influence of dropout rates) we furthermore trained 10 models for every dropout rate $p \in [0.1, 0.2, \ldots, 0.9]$ for the Mnist, Cifar10 and Traffic case studies. 

Counting atomic models individually, this procedure required the training of 246 atomic DNNs,
and the inference of more than 3 billion atomic DNN predictions\footnote{Predictions were cached, such that for the evaluation of different sample sizes and different numbers
of atomic models, previous predictions could be re-used.}.
Due to the high workload, the training and prediction processes were distributed on two different workstations using Windows or Ubuntu and three different GPUs (the Ubuntu workstation had two GPUs). 
\tool\cite{Weiss2021a} was used for training, prediction and uncertainty quantification.

\subsection{Results}
  \afterpage{%

\begingroup
\setlength{\tabcolsep}{6pt} %
\renewcommand{\arraystretch}{1} %

\newcommand{\vertimodeltype}[1]{\begin{tabular}{@{}c@{}}\rotatebox[origin=c]{90}{\parbox{1cm}{\centering #1}}\end{tabular}}
\def \spacebetweenbigtrows {\vspace{.6cm}}
\def \spacebetweenquantifiers {\vspace{.15cm}}

\begin{table*}[]
\scriptsize
\resizebox{\textwidth}{!}{
\begin{tabular}{lllrrrrrrrrrrrrrrrrrr}
\toprule
                             &                              &            & \multicolumn{9}{c}{Nominal (regular test data)}                                                                                                                                                                                                                                    & \multicolumn{9}{c}{Out of Distribution (corrupted test data)}\\
\cmidrule(r){4-12}
\cmidrule(r){13-21}
                             &                              &            &  & \multicolumn{4}{c}{$\epsilon$ = 0.01}                                                                        & \multicolumn{4}{c}{$\epsilon$ = 0.1}                                                                         &  & \multicolumn{4}{c}{$\epsilon$ = 0.01}                                                                        & \multicolumn{4}{c}{$\epsilon$ = 0.1}                                                                         \\
\cmidrule(r){5-8}
\cmidrule(r){9-12}
\cmidrule(r){14-17}
\cmidrule(r){18-21}

& \multicolumn{2}{c}{Technique}                & \multicolumn{1}{r}{$\scriptstyle ACC$}        & \multicolumn{1}{r}{$\overline{\scriptstyle ACC}$} & \multicolumn{1}{r}{$\Delta_u$} & \multicolumn{1}{r}{$S_1$} & S-C & \multicolumn{1}{r}{$\overline{\scriptstyle ACC}$} & \multicolumn{1}{r}{$\Delta_u$} & \multicolumn{1}{r}{$S_1$} & S-C & \multicolumn{1}{r}{$\scriptstyle ACC$}         & \multicolumn{1}{r}{$\overline{\scriptstyle ACC}$} & \multicolumn{1}{r}{$\Delta_u$} & \multicolumn{1}{r}{$S_1$} & S-C & \multicolumn{1}{r}{$\overline{\scriptstyle ACC}$} & \multicolumn{1}{r}{$\Delta_u$} & \multicolumn{1}{r}{$S_1$} & S-C\\
\midrule\ \\
\multirow{15}{*}{\vertimodeltype{cifar10}}    & \multirow{3}{*}{\vertimodeltype{Point Pred.}} & SM         & 0.82                             & 0.83                                 & 0.98                           & 0.90                      & -0.96                     & 0.89                                 & 0.83                           & 0.86                      & -0.96                     & 0.82                             & 0.83                                 & 0.98                           & 0.90                      & -0.90                     & 0.89                                 & 0.82                           & 0.86                      & -0.91                     \\
                             &                              & PCS        & 0.82                             & 0.83                                 & 0.98                           & 0.90                      & -0.96                     & 0.89                                 & 0.82                           & 0.85                      & -0.96                     & 0.82                             & 0.83                                 & 0.98                           & 0.90                      & -0.90                     & 0.89                                 & 0.82                           & 0.86                      & -0.89                     \\
                             &                              & SME        & 0.82                             & 0.83                                 & 0.98                           & 0.90                      & -0.96                     & 0.89                                 & 0.82                           & 0.85                      & -0.96                     & 0.82                             & 0.83                                 & 0.98                           & 0.90                      & -0.90                     & 0.89                                 & 0.82                           & 0.86                      & -0.90\spacebetweenquantifiers                     \\
                             & \multirow{4}{*}{\vertimodeltype{MC- Dropout}}  & VR         & 0.82                             & 0.83                                 & 0.98                           & 0.90                      & -0.93                     & 0.90                                 & 0.83                           & 0.86                      & -0.88                     & 0.83                             & 0.84                                 & 0.98                           & 0.90                      & -0.89                     & 0.90                                 & 0.83                           & 0.86                      & -0.87                     \\
                             &                              & PE         & 0.82                             & 0.83                                 & 0.98                           & 0.90                      & -0.95                     & 0.89                                 & 0.84                           & 0.86                      & -0.95                     & 0.83                             & 0.83                                 & 0.99                           & 0.90                      & -0.90                     & 0.89                                 & 0.84                           & 0.87                      & -0.91                     \\
                             &                              & MI         & 0.82                             & 0.83                                 & 0.98                           & 0.90                      & -0.95                     & 0.88                                 & 0.85                           & 0.86                      & -0.94                     & 0.83                             & 0.84                                 & 0.98                           & 0.90                      & -0.91                     & 0.89                                 & 0.84                           & 0.86                      & -0.91                     \\
                             &                              & MS         & 0.82                             & 0.83                                 & 0.98                           & 0.90                      & -0.96                     & 0.89                                 & 0.83                           & 0.86                      & -0.96                     & 0.83                             & 0.84                                 & 0.98                           & 0.90                      & -0.90                     & 0.90                                 & 0.83                           & 0.86                      & -0.90\spacebetweenquantifiers                    \\
                             & \multirow{4}{*}{\vertimodeltype{Ensem- ble}}    & VR         & 0.86                             & 0.88                                 & 0.97                           & 0.92                      & -0.90                     & 0.94                                 & 0.83                           & 0.88                      & -0.90                     & 0.86                             & 0.88                                 & 0.97                           & 0.92                      & -0.91                     & 0.93                                 & 0.83                           & 0.88                      & -0.91                     \\
                             &                              & PE         & 0.86                             & 0.87                                 & 0.97                           & 0.92                      & -0.95                     & 0.93                                 & 0.83                           & 0.88                      & -0.95                     & 0.86                             & 0.88                                 & 0.98                           & 0.92                      & -0.94                     & 0.92                                 & 0.84                           & 0.88                      & -0.92                     \\
                             &                              & MI         & 0.86                             & 0.87                                 & 0.98                           & 0.92                      & -0.94                     & 0.93                                 & 0.83                           & 0.88                      & -0.92                     & 0.86                             & 0.88                                 & 0.98                           & 0.92                      & -0.93                     & 0.92                                 & 0.84                           & 0.88                      & -0.92                     \\
                             &                              & MS         & 0.86                             & 0.88                                 & 0.97                           & 0.92                      & -0.96                     & 0.94                                 & 0.83                           & 0.88                      & -0.96                     & 0.87                             & 0.88                                 & 0.97                           & 0.92                      & -0.95                     & 0.93                                 & 0.83                           & 0.88                      & -0.94\spacebetweenquantifiers                     \\
                             & \multirow{4}{*}{\vertimodeltype{Flip- out}}              & VR         & 0.69                             & 0.71                                 & 0.98                           & 0.82                      & -0.62                     & 0.77                                 & 0.82                           & 0.80                      & -0.54                     & 0.71                             & 0.72                                 & 0.98                           & 0.83                      & -0.56                     & 0.77                                 & 0.82                           & 0.80                      & -0.52                     \\
                             &                              & PE         & 0.69                             & 0.70                                 & 0.98                           & 0.82                      & -0.65                     & 0.76                                 & 0.82                           & 0.79                      & -0.65                     & 0.71                             & 0.72                                 & 0.98                           & 0.83                      & -0.58                     & 0.77                                 & 0.82                           & 0.79                      & -0.59                     \\
                             &                              & MI         & 0.69                             & 0.70                                 & 0.99                           & 0.82                      & -0.66                     & 0.73                                 & 0.86                           & 0.79                      & -0.65                     & 0.71                             & 0.71                                 & 0.99                           & 0.83                      & -0.57                     & 0.73                                 & 0.86                           & 0.79                      & -0.54                     \\
                             &                              & MS         & 0.70                             & 0.71                                 & 0.97                           & 0.82                      & -0.65                     & 0.77                                 & 0.81                           & 0.79                      & -0.66                     & 0.71                             & 0.72                                 & 0.97                           & 0.83                      & -0.60                     & 0.77                                 & 0.82                           & 0.80                      & -0.64\spacebetweenbigtrows                      \\
\multirow{15}{*}{\vertimodeltype{mnist}}      & \multirow{3}{*}{\vertimodeltype{Point Pred.}} & SM         & 0.97                             & 0.97                                 & 0.98                           & 0.98                      & -0.83                     & 1.00                                 & 0.88                           & 0.93                      & -0.88                     & 0.76                             & 0.85                                 & 0.82                           & 0.83                      & -0.47                     & 0.96                                 & 0.49                           & 0.65                      & -0.74                     \\
                             &                              & PCS        & 0.97                             & 0.97                                 & 0.98                           & 0.98                      & -0.83                     & 1.00                                 & 0.88                           & 0.94                      & -0.86                     & 0.76                             & 0.82                                 & 0.88                           & 0.85                      & -0.49                     & 0.96                                 & 0.50                           & 0.66                      & -0.73                     \\
                             &                              & SME        & 0.97                             & 0.97                                 & 0.99                           & 0.98                      & -0.83                     & 1.00                                 & 0.88                           & 0.93                      & -0.88                     & 0.76                             & 0.86                                 & 0.79                           & 0.82                      & -0.47                     & 0.96                                 & 0.48                           & 0.64                      & -0.75\spacebetweenquantifiers                     \\
                             & \multirow{4}{*}{\vertimodeltype{MC- Dropout}}  & VR         & 0.97                             & 0.97                                 & 0.98                           & 0.98                      & -0.80                     & 1.00                                 & 0.87                           & 0.93                      & -0.88                     & 0.75                             & 0.84                                 & 0.85                           & 0.84                      & -0.37                     & 0.96                                 & 0.52                           & 0.68                      & -0.70                     \\
                             &                              & PE         & 0.97                             & 0.97                                 & 0.99                           & 0.98                      & -0.83                     & 1.00                                 & 0.88                           & 0.93                      & -0.87                     & 0.75                             & 0.86                                 & 0.79                           & 0.82                      & -0.45                     & 0.96                                 & 0.47                           & 0.63                      & -0.76                     \\
                             &                              & MI         & 0.97                             & 0.97                                 & 0.99                           & 0.98                      & -0.83                     & 0.99                                 & 0.87                           & 0.93                      & -0.89                     & 0.75                             & 0.81                                 & 0.86                           & 0.83                      & -0.54                     & 0.92                                 & 0.56                           & 0.70                      & -0.65                     \\
                             &                              & MS         & 0.96                             & 0.97                                 & 0.98                           & 0.98                      & -0.83                     & 1.00                                 & 0.88                           & 0.93                      & -0.88                     & 0.75                             & 0.86                                 & 0.80                           & 0.83                      & -0.39                     & 0.96                                 & 0.49                           & 0.65                      & -0.79\spacebetweenquantifiers                     \\
                             & \multirow{4}{*}{\vertimodeltype{Ensem- ble}}    & VR         & 0.97                             & 0.97                                 & 0.98                           & 0.98                      & -0.85                     & 0.98                                 & 0.95                           & 0.97                      & -0.84                     & 0.74                             & 0.85                                 & 0.80                           & 0.82                      & -0.37                     & 0.91                                 & 0.65                           & 0.76                      & -0.76                     \\
                             &                              & PE         & 0.97                             & 0.97                                 & 0.99                           & 0.98                      & -0.88                     & 0.99                                 & 0.88                           & 0.93                      & -0.91                     & 0.74                             & 0.86                                 & 0.77                           & 0.81                      & -0.50                     & 0.96                                 & 0.45                           & 0.62                      & -0.87                     \\
                             &                              & MI         & 0.97                             & 0.97                                 & 0.98                           & 0.98                      & -0.90                     & 1.00                                 & 0.88                           & 0.93                      & -0.94                     & 0.74                             & 0.86                                 & 0.72                           & 0.78                      & -0.73                     & 0.98                                 & 0.42                           & 0.59                      & -0.79                     \\
                             &                              & MS         & 0.97                             & 0.97                                 & 0.98                           & 0.98                      & -0.88                     & 1.00                                 & 0.88                           & 0.93                      & -0.90                     & 0.75                             & 0.86                                 & 0.78                           & 0.82                      & -0.55                     & 0.96                                 & 0.46                           & 0.63                      & -0.85\spacebetweenquantifiers                     \\
                             & \multirow{4}{*}{\vertimodeltype{Flip- out}}     & VR         & 0.94                             & 0.95                                 & 0.98                           & 0.97                      & -0.93                     & 0.98                                 & 0.87                           & 0.92                      & -0.96                     & 0.71                             & 0.79                                 & 0.84                           & 0.82                      & -0.80                     & 0.91                                 & 0.54                           & 0.68                      & -0.88                     \\
                             &                              & PE         & 0.94                             & 0.95                                 & 0.98                           & 0.97                      & -0.94                     & 0.98                                 & 0.87                           & 0.92                      & -0.95                     & 0.71                             & 0.82                                 & 0.78                           & 0.80                      & -0.87                     & 0.92                                 & 0.48                           & 0.63                      & -0.93                     \\
                             &                              & MI         & 0.94                             & 0.95                                 & 0.98                           & 0.97                      & -0.94                     & 0.98                                 & 0.87                           & 0.92                      & -0.95                     & 0.71                             & 0.77                                 & 0.82                           & 0.80                      & -0.89                     & 0.86                                 & 0.57                           & 0.69                      & -0.86                     \\
                             &                              & MS         & 0.94                             & 0.95                                 & 0.98                           & 0.97                      & -0.93                     & 0.98                                 & 0.87                           & 0.92                      & -0.96                     & 0.71                             & 0.81                                 & 0.82                           & 0.81                      & -0.85                     & 0.93                                 & 0.51                           & 0.66                      & -0.94\spacebetweenbigtrows                      \\
\multirow{15}{*}{\vertimodeltype{traffic}}    & \multirow{3}{*}{\vertimodeltype{Point Pred.}} & SM         & 0.79                             & 0.88                                 & 0.89                           & 0.89                      & -0.26                     & 0.96                                 & 0.74                           & 0.84                      & -0.30                     & 0.78                             & 0.88                                 & 0.87                           & 0.87                      & -0.56                     & 0.97                                 & 0.70                           & 0.81                      & -0.36                     \\
                             &                              & PCS        & 0.79                             & 0.88                                 & 0.89                           & 0.89                      & -0.34                     & 0.97                                 & 0.74                           & 0.84                      & -0.28                     & 0.78                             & 0.88                                 & 0.87                           & 0.87                      & -0.54                     & 0.97                                 & 0.69                           & 0.81                      & -0.36                     \\
                             &                              & SME        & 0.79                             & 0.88                                 & 0.89                           & 0.89                      & -0.40                     & 0.98                                 & 0.71                           & 0.82                      & -0.30                     & 0.78                             & 0.88                                 & 0.87                           & 0.87                      & -0.62                     & 0.98                                 & 0.66                           & 0.79                      & -0.37\spacebetweenquantifiers                     \\
                             & \multirow{4}{*}{\vertimodeltype{MC- Dropout}}  & VR         & 0.79                             & 0.89                                 & 0.88                           & 0.88                      & -0.53                     & 0.98                                 & 0.69                           & 0.81                      & -0.69                     & 0.78                             & 0.88                                 & 0.87                           & 0.87                      & -0.65                     & 0.98                                 & 0.65                           & 0.78                      & -0.77                     \\
                             &                              & PE         & 0.79                             & 0.88                                 & 0.89                           & 0.88                      & -0.52                     & 0.98                                 & 0.69                           & 0.81                      & -0.64                     & 0.78                             & 0.88                                 & 0.86                           & 0.87                      & -0.57                     & 0.98                                 & 0.65                           & 0.78                      & -0.67                     \\
                             &                              & MI         & 0.79                             & 0.88                                 & 0.89                           & 0.88                      & -0.51                     & 0.98                                 & 0.69                           & 0.81                      & -0.72                     & 0.78                             & 0.87                                 & 0.88                           & 0.87                      & -0.65                     & 0.98                                 & 0.65                           & 0.78                      & -0.81                     \\
                             &                              & MS         & 0.80                             & 0.88                                 & 0.88                           & 0.88                      & -0.50                     & 0.98                                 & 0.69                           & 0.81                      & -0.62                     & 0.78                             & 0.88                                 & 0.86                           & 0.87                      & -0.58                     & 0.98                                 & 0.65                           & 0.78                      & -0.66\spacebetweenquantifiers                     \\
                             & \multirow{4}{*}{\vertimodeltype{Ensem- ble}}    & VR         & 0.81                             & 0.94                                 & 0.84                           & 0.89                      & -0.67                     & 0.97                                 & 0.77                           & 0.86                      & -0.91                     & 0.80                             & 0.94                                 & 0.82                           & 0.87                      & -0.62                     & 0.97                                 & 0.74                           & 0.84                      & -0.92                     \\
                             &                              & PE         & 0.81                             & 0.93                                 & 0.86                           & 0.89                      & -0.47                     & 0.99                                 & 0.68                           & 0.81                      & -0.70                     & 0.80                             & 0.93                                 & 0.83                           & 0.88                      & -0.51                     & 0.99                                 & 0.63                           & 0.77                      & -0.70                     \\
                             &                              & MI         & 0.81                             & 0.94                                 & 0.84                           & 0.89                      & -0.76                     & 0.99                                 & 0.68                           & 0.81                      & -0.70                     & 0.80                             & 0.94                                 & 0.82                           & 0.87                      & -0.74                     & 0.99                                 & 0.63                           & 0.77                      & -0.73                     \\
                             &                              & MS         & 0.81                             & 0.94                                 & 0.84                           & 0.89                      & -0.62                     & 0.99                                 & 0.68                           & 0.81                      & -0.71                     & 0.80                             & 0.94                                 & 0.82                           & 0.88                      & -0.61                     & 0.99                                 & 0.63                           & 0.77                      & -0.69\spacebetweenquantifiers                     \\
                             & \multirow{4}{*}{\vertimodeltype{Flip- out}}     & VR         & 0.78                             & 0.87                                 & 0.87                           & 0.87                      & -0.89                     & 0.97                                 & 0.67                           & 0.79                      & -0.95                     & 0.75                             & 0.85                                 & 0.86                           & 0.85                      & -0.89                     & 0.98                                 & 0.62                           & 0.76                      & -0.95                     \\
                             &                              & PE         & 0.78                             & 0.84                                 & 0.90                           & 0.87                      & -0.90                     & 0.97                                 & 0.68                           & 0.80                      & -0.93                     & 0.75                             & 0.84                                 & 0.87                           & 0.85                      & -0.91                     & 0.97                                 & 0.63                           & 0.76                      & -0.93                     \\
                             &                              & MI         & 0.78                             & 0.85                                 & 0.90                           & 0.87                      & -0.90                     & 0.97                                 & 0.68                           & 0.80                      & -0.93                     & 0.75                             & 0.83                                 & 0.88                           & 0.86                      & -0.90                     & 0.97                                 & 0.63                           & 0.77                      & -0.92                     \\
                             &                              & MS         & 0.77                             & 0.86                                 & 0.88                           & 0.87                      & -0.90                     & 0.97                                 & 0.68                           & 0.80                      & -0.97                     & 0.75                             & 0.84                                 & 0.87                           & 0.85                      & -0.91                     & 0.97                                 & 0.63                           & 0.76                      & -0.96\spacebetweenbigtrows                      \\
\multirow{7}{*}{\vertimodeltype{Efficientnet}} & \multirow{3}{*}{\vertimodeltype{Point Pred.}} & SM         & 0.74                             & 0.77                                 & 0.95                           & 0.85                      & n.a.  & 0.84                                 & 0.79                           & 0.81                      & n.a.  & 0.50                             & 0.61                                 & 0.79                           & 0.69                      & n.a.  & 0.75                                 & 0.54                           & 0.63                      & n.a.  \\
                             &                              & PCS        & 0.74                             & 0.76                                 & 0.97                           & 0.85                      & n.a.  & 0.84                                 & 0.79                           & 0.81                      & n.a.  & 0.50                             & 0.55                                 & 0.89                           & 0.68                      & n.a.  & 0.72                                 & 0.57                           & 0.64                      & n.a.  \\
                             &                              & SME        & 0.74                             & 0.76                                 & 0.96                           & 0.85                      & n.a.  & 0.83                                 & 0.80                           & 0.81                      & n.a.  & 0.50                             & 0.60                                 & 0.80                           & 0.68                      & n.a.  & 0.73                                 & 0.55                           & 0.63                      & n.a.\spacebetweenquantifiers  \\
                             & \multirow{4}{*}{\vertimodeltype{MC- Dropout}}  & VR         & 0.74                             & 0.76                                 & 0.96                           & 0.85                      & -0.64                     & 0.84                                 & 0.78                           & 0.81                      & -0.83                     & 0.50                             & 0.56                                 & 0.87                           & 0.68                      & -0.86                     & 0.71                                 & 0.57                           & 0.63                      & -0.95                     \\
                             &                              & PE         & 0.74                             & 0.76                                 & 0.96                           & 0.85                      & -0.33                     & 0.83                                 & 0.79                           & 0.81                      & -0.60                     & 0.50                             & 0.61                                 & 0.78                           & 0.69                      & -0.83                     & 0.75                                 & 0.53                           & 0.62                      & -0.86                     \\
                             &                              & MI         & 0.74                             & 0.75                                 & 0.98                           & 0.85                      & -0.72                     & 0.82                                 & 0.81                           & 0.81                      & -0.83                     & 0.50                             & 0.52                                 & 0.93                           & 0.67                      & -0.86                     & 0.67                                 & 0.58                           & 0.62                      & -0.93                     \\
                             &                              & MS         & 0.74                             & 0.77                                 & 0.96                           & 0.85                      & -0.31                     & 0.85                                 & 0.78                           & 0.81                      & -0.35                     & 0.50                             & 0.61                                 & 0.79                           & 0.69                      & -0.80                     & 0.76                                 & 0.52                           & 0.62                      & -0.75                    \\ \bottomrule
\end{tabular}
}

\caption{Overview of supervision capabilities of different techniques, i.e., model types and quantifier combinations for different thresholds.}
\label{tab:big_table}
\end{table*}

\endgroup

    \clearpage
  }
We organize the analysis of the results obtained in our experiments by research question.

\subsubsection{RQ1 (Effectiveness)}

An overview of our results is provided in \autoref{tab:big_table}. 
Due to space constraints, the results for $\epsilon=0.05$ are omitted in the table and the values for the 8 different Imagenet models are averaged. 
The full set of results can be found in the online replication package.

Our results suggest that all supervisors lead to supervised accuracies $\overline{ACC}$ which are at least as high, but typically much higher than the accuracy $ACC$ of the unsupervised model.
Thus,  supervisors are effective. 
The effectiveness is particularly strong on the OOD datasets.
For example, on Mnist, where the unsupervised point predictor has an accuracy of 76\%, 
the supervised accuracy at $\epsilon=0.1$, 
is above 95\% with most supervisors. 
In other words, a DLS using an unsupervised model will experience six times more faulty predictions than the supervised one.
Also notable are the results on the nominal Mnist dataset, where even simple supervisors based on  point predictors turn an unsupervised accuracy of 97\% (at $\epsilon = 0.1$) into that of a nearly perfect predictor 
while still accepting around 88\% of the inputs.

It is worth mentioning that when training the Flipout model for the traffic case study, we faced randomly occurring problems that prevented the model from even converging to a state allowing useful predictions, let alone uncertainty quantification.
We ran our evaluation script multiple times and report one successful training run in \autoref{tab:big_table}, to allow fair comparison with the other uncertainty-aware DNNs.
To investigate and discuss the training problem we faced, we conducted an additional set of experiments where we repeatedly train Flipout models under different hyperparameters. The results are presented in Appendix \ref{sec:appendix_traffic_bnn}.
Summarized, we find that Flipout models often need a long \emph{burn-in} phase, before training accuracy starts increasing - sometimes not even reaching this improving phase within the first 200 epochs.
Aiming to fast-forward through this phase by using a higher learning rate consistently led to \textit{Not-A-Number (NaN)} errors, likely caused by exploding gradients.
Implementing only few of the convolutional and dense layers as Flipout layers (leaving the other ones regular PPNN layers) reliably solved these problems.
With regards to RQ1, for \emph{successfully} trained Flipout models, the quantified uncertainty clearly showed the capability to perform useful UQ, i.e., to lead to a supervised accuracy which is higher than the unsupervised accuracy. 

\vspace{0.5cm}
\begin{tcolorbox}
\textbf{Summary (RQ1)}: \textit{All tested supervisors are, in general, effective at increasing the supervised accuracy compared to the unsupervised accuracy.} 
\end{tcolorbox}

\subsubsection{RQ2 (Comparison)}
If we look at the $S_1$-Score in \autoref{tab:big_table}, it is apparent that there is no uncertainty quantifier that outperforms all the other ones on every subject/dataset and for every threshold ($\epsilon$). 
To allow for an overall comparison of the quantifiers, we computed the average ranks of the quantifiers  when ordered by $S_1$-Score. Results are shown in \autoref{tab:rank_order}, where $N$ indicates the number of data points on which average ranks have been computed. While there is no absolute dominant supervisor, i.e., the ideal choice of supervisor remains problem-dependent,
Ensembles can be considered the overall best performing supervisors (in line with existing literature\cite{Ovadia2019}), because when they do not have the lowest rank, they still have quite low-rank values. Actually, 
even without supervision Ensembles often achieve higher accuracy than Point Predictors and MC-Dropout-based models.
Interestingly, in most cases the sophisticated quantifiers PE and MI, grounded on information theory,
do not perform better and often perform  worse than the simpler quantifiers VR and AS. 
Despite the theoretical disadvantages of Point Predictors, in our experiments,
this simple approach outperformed the theoretically well-founded MC-Dropout. So, in practical cases as those considered in our experiments, Point Predictors may represent a good trade-off between performance and computational cost.

While there is no best technique, in our experiments there clearly is a worst one. 
The models with flipout layers perform worse than all other investigated models with equivalent architectures but non-probabilistic weights for all case studies and with all quantifiers. 
It is worth noting here that this finding does not \emph{generally} mean that flipout-based PW-BNN are worse than other architectures, but that we should not just create PW-BNN by replacing deterministic layers with probabilistic layers and keeping all other model properties unchanged, which is sometimes suggested in tutorials on BNN implementation~\cite{Dillon2019}.
Due to the generally lower performance and unstable training process of Flipout models, we focus our answer to RQ3 and RQ4 (hyperparameter sensitivity) around Deep Ensembles and MC-Dropout, putting less emphasis on Flipout models. 

\vspace{0.5cm}
\begin{tcolorbox}
\textbf{Summary (RQ2)}: \textit{There is no dominant supervisor, i.e., no supervisor which performs best for every test subject, data source and threshold. Ensembles, in particular using VR as quantifier, are ranked generally well across subjects and thresholds, while Point Predictors offer a valuable trade off between performance and execution cost.
Without careful tuning, Flipout based PW-BNN perform drastically worse than the other approaches, even compared to the simple PPNN using an equivalent architecture.} 
\end{tcolorbox}

\newcommand{\verti}[1]{\begin{tabular}{@{}c@{}}\rotatebox[origin=c]{90}{\parbox{1cm}{\centering #1}}\end{tabular}}
\def \spacebetweenrankrows {0.2cm}
\begin{table}[t]
\centering
\begin{tabular}{@{}llcccccccc@{}}
\toprule
                                      &             &  \multicolumn{3}{c}{Per Subject}                                                                                          &&   \multicolumn{2}{c}{Overall}             \\
                                      &              & mnist                                    & cifar10                                  & traffic                                   &&  Full Studies                                & Pre-Trained                                                   \\
\multicolumn{2}{c}{Technique}                       & N=6                                & N=6                                & N=6                                &&  N=18                                 & N=48           \\ 
\midrule
\multirow{3}{*}{\verti{Point Pred.}}   & SM     & 4.67     & 9.83     & 4.17     &&  6.22     & 3.56             \\
                                      & PCS         & 3.5             & 10.5             & 4.67             &&  6.22             & 3.5                       \\
                                      & SME     & 5.67 & 9.67 & 4.67 &&  6.67 & 4.42  \vspace{\spacebetweenrankrows}               \\
\multirow{4}{*}{\verti{MC- Dropout}}      & VR  & 6.0       & 6.5       & 9.5       &&  7.33       & 3.55             \\
                                                   & PE  & 7.5    & 6.5    & 7.58    &&  7.19    & 4.5             \\
                                                   & MI & 5.0       & 6.83       & 6.5       &&  6.11       & 5.12             \\
                                                   & AS & 7.17         & 6.17         & 7.42         &&  6.92         & 3.34   \vspace{\spacebetweenrankrows}                \\
\multirow{4}{*}{\verti{Ensemble}}       & VR  & 3.83         & 3.0         & 1.67         &&  2.83         & n.a.                                                 \\
                                                   & PE  & 9.83      & 1.83      & 6.58      &&  6.08      & n.a.                                                 \\
                                                   & MI & 11.83         & 2.67         & 8.33         &&  7.61         & n.a.                                                 \\
                                                   & AS & 9.83           & 2.5           & 6.58           &&  6.31           & n.a.                   \vspace{\spacebetweenrankrows}                \\                  
\multirow{4}{*}{\verti{Flipout}}       & VR  & 9.83         & 12.33         & 14.33         &&  12.17         & n.a.                                                 \\
                                                   & PE  & 12.83      & 14.17      & 11.83      &&  12.94      & n.a.                                                 \\
                                                   & MI & 11.5         & 14.83         & 13.0         &&  13.11         & n.a.                                                 \\
                                                   & AS & 11.0           & 12.67           & 13.17           &&  12.28           & n.a.                                     
\end{tabular}
\caption{Rank-Order Analysis: $S_1$-Score ranks, averaged over $\epsilon\in\{0.01, 0.05, 0.1\}$, nominal and OOD datasets}
\label{tab:rank_order}
\end{table}

\subsubsection{RQ3 (Sample size)}
\def \sssubfigurewidth {.45\linewidth}
\def \ssimgwidth {\linewidth}

\begin{figure}
\centering
\begin{subfigure}{\sssubfigurewidth}
  \centering
  \includegraphics[width=\ssimgwidth]{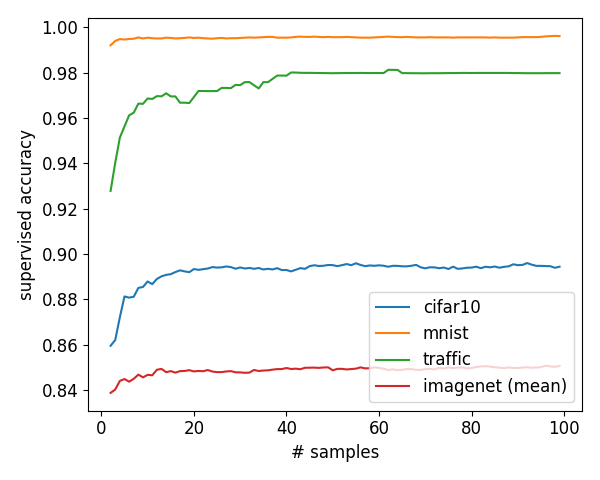}
  \caption{MC-Dropout}
  \label{fig:ss_influence_mcdropout}
\end{subfigure}%
\begin{subfigure}{\sssubfigurewidth}
  \centering
  \includegraphics[width=\ssimgwidth]{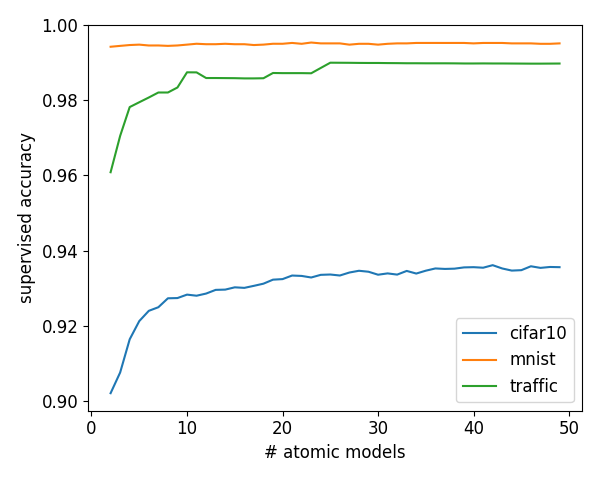}
  \caption{Deep Ensemble}
  \label{fig:ss_influence_ensemble}
\end{subfigure}

\caption{Influence of sample size in MC-Dropout (\subref{fig:ss_influence_mcdropout}) and number of atomic models in Ensemble (\subref{fig:ss_influence_ensemble}) on supervised accuracy $\overline{acc}$;
values taken with $\epsilon=0.1$ and MS quantifier on nominal dataset}
\label{fig:ss_influence}
\end{figure}
We find that for both MC-Dropout and Ensembles the relatively low number of 20 samples
is already sufficient to get a similar supervised accuracy as with a much higher number of samples.
This is shown in \autoref{fig:ss_influence} for the MS quantifier and $\epsilon = 0.1$. 
20 samples, while still higher than what's recommended in related literature~\cite{Ovadia2019},
is probably small enough to be used in practice in many applications.
The other quantifiers behave similarly, with one notable exception, which is the VR quantifier: VR can only take a finite set of discrete values, which can be easily shown to be equal to the number of samples. 
Hence, a low sample size makes it impossible to set thresholds well fit to the target $\epsilon$, because thresholds are correspondingly also discrete and limited to the number of samples. So, to achieve the target FPR $\epsilon$ precisely, we might need substantially more than 20 samples with VR.

\vspace{0.5cm}
\begin{tcolorbox}
\textbf{Summary (RQ3)}: \textit{A few ($\sim 20$) samples are  enough to get good supervision results with most quantifiers (VR represents an exception, due to the discretization of the values it can take).} 
\end{tcolorbox}

\subsubsection{RQ4 (Sensitivity)}
\afterpage{
\def \hmsubfigurewidth {.4\linewidth}

\def \hmimgwidth {6cm}
\def \hmimgheight {6.3cm}

\begin{figure}
\centering
\begin{subfigure}{\hmsubfigurewidth}
  \centering
  \includegraphics[
  height=\hmimgheight
  ]{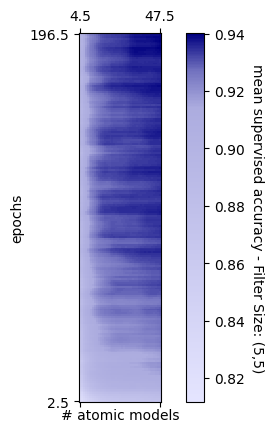}
  \caption{Supervised Accuracy Average}
  \label{fig:heatmap_mean}
\end{subfigure}%
\hspace{1cm}
\begin{subfigure}{\hmsubfigurewidth}
   \centering
  \includegraphics[
  height=\hmimgheight
  ]{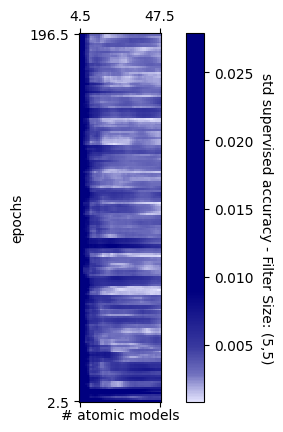}
  \caption{Supervised Accuracy Std. Dev.}
  \label{fig:heatmap_std}
\end{subfigure}

\caption{Average (\subref{fig:ss_influence_mcdropout}) and standard deviation (\subref{fig:ss_influence_ensemble})
of the Ensemble's supervised accuracy $\overline{acc}$ with $\epsilon=0.1$ on OOD data, computed for the Traffic subject inside a 5x5 neighborhood of each configuration}
\label{fig:heatmaps}
\end{figure}
}
The number of training epochs and the number of samples/atomic models can be visualized as a 200 (number of epochs) by 100 (number of samples) or 50 (number of atomic models) grid. 
To assess the sensitivity of supervisors, we calculate the standard deviation (\textit{std}) of $\overline{ACC}$ within a 5x5 filter applied to this grid. In this way, we account for the variability of the supervised accuracy in a 5x5 neighbourhood of each hyperparameter configuration.
The higher the std, the higher the sensitivity to small hyperparameter changes in the neighbourhood.
We also calculate the average $\overline{ACC}$ in such 5x5 neighbourhood.
The result is a pair of heatmaps, an example of which is shown in \autoref{fig:heatmaps}. 
This figure suggests that hyperparameter sensitivity negatively correlates with $\overline{ACC}$: 
Hyperparameters leading to low accuracy (bright colors in \autoref{fig:heatmap_mean}) typically show high sensitivity to hyperparameter changes (dark colors in \autoref{fig:heatmap_std}).
We do indeed observe this negative correlation for all case studies.
The values of Pearson correlation, shown in the \emph{S-C} (Sensitivity-Correlation) columns of \autoref{tab:big_table}, are strongly negative in most cases, with $p$-value $< 0.05$ in 457 out of 462 cases.
In low accuracy cases, small changes to number of samples and number of epochs have a high effect on accuracy.

\vspace{0.5cm}
\begin{tcolorbox}
\textbf{Summary (RQ4)}: \textit{For a given supervisor and model, for what concerns the number of training epochs and samples used for quantification, the higher the supervised accuracy, the lower the model's sensitivity to small hyperparameter changes.} 
\end{tcolorbox}

\subsubsection{RQ5 (Dropout Rate)}
\afterpage{

\def \dropsubfigurewidth {0.33\linewidth}
\def \dropimgwidth {1.1\linewidth}

\begin{figure}[t]
\centering
\begin{subfigure}{\dropsubfigurewidth}
  \centering
  \includegraphics[width=\dropimgwidth]{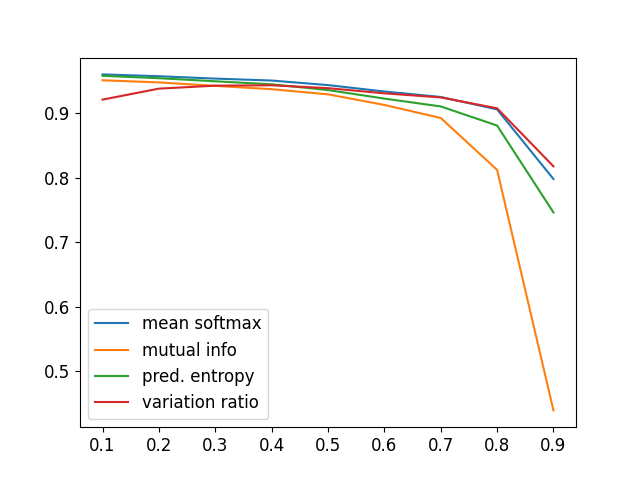}
  \caption{MNIST Nominal}
  \label{fig:drop_mnist}
\end{subfigure}%
\begin{subfigure}{\dropsubfigurewidth}
  \centering
  \includegraphics[width=\dropimgwidth]{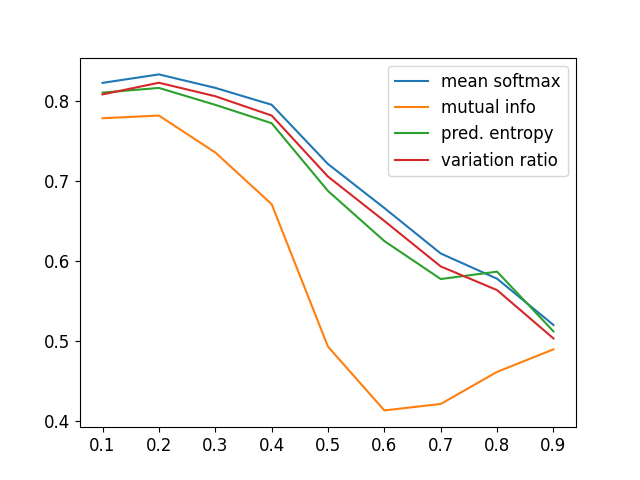}
  \caption{CIFAR-10 Nominal}
  \label{fig:drop_cifar}
\end{subfigure}
\begin{subfigure}{\dropsubfigurewidth}
  \centering
  \includegraphics[width=\dropimgwidth]{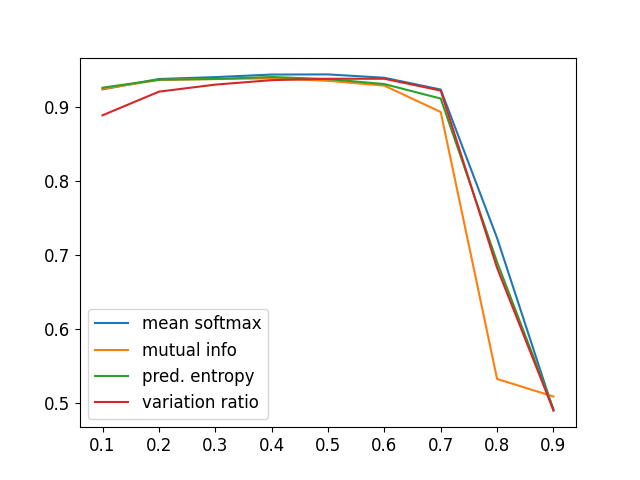}
  \caption{TRAFFIC Nominal}
  \label{fig:drop_pr}
\end{subfigure}

\begin{subfigure}{\dropsubfigurewidth}
  \centering
  \includegraphics[width=\dropimgwidth]{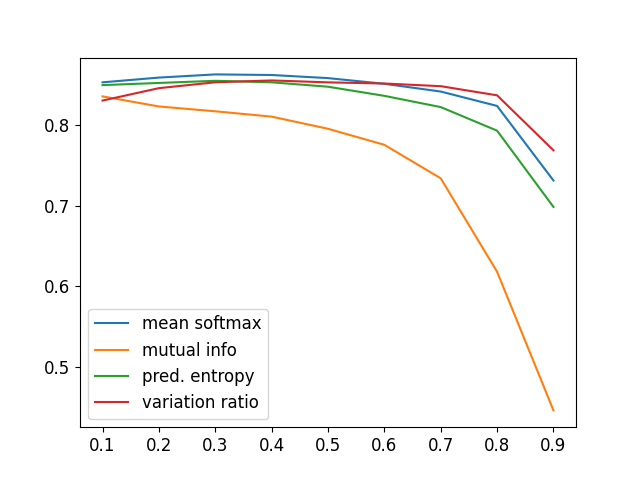}
  \caption{MNIST OOD}
  \label{fig:drop_mnist_ood}
\end{subfigure}%
\begin{subfigure}{\dropsubfigurewidth}
  \centering
  \includegraphics[width=\dropimgwidth]{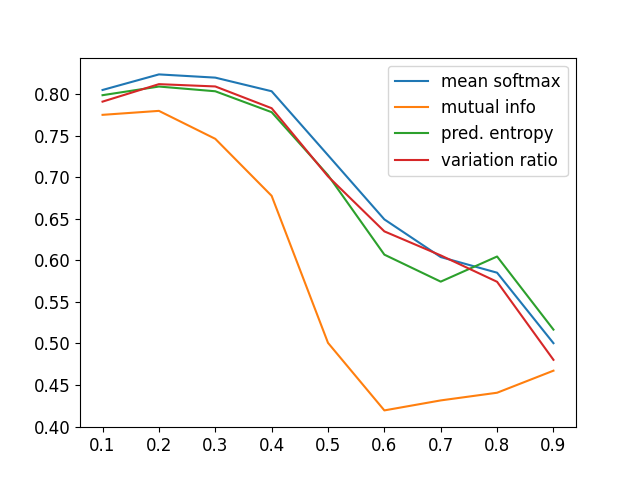}
  \caption{CIFAR-10 OOD}
  \label{fig:drop_cifar_ood}
\end{subfigure}
\begin{subfigure}{\dropsubfigurewidth}
  \centering
  \includegraphics[width=\dropimgwidth]{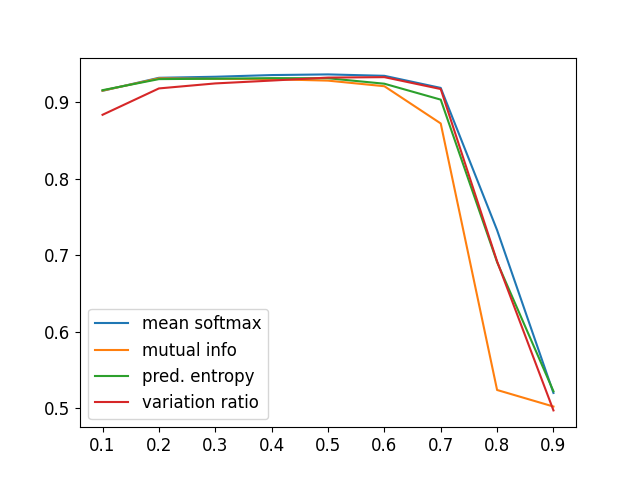}
  \caption{TRAFFIC OOD}
  \label{fig:drop_pr_ood}
\end{subfigure}

\caption{Influence of Dropout Rate on UQ: \changed{AVGPR} at increasing dropout rates}
\label{fig:dropout_rates}
\end{figure}
}

To measure the impact of the choice of dropout rates $p$ in MC-Dropout, we tested different dropout rates $p \in \{0.1, 0.2, \ldots, 0.9\}$ for the Mnist, Cifar10 and Traffic case studies. 
To perform a threshold-independent analysis, UQ capabilities have been measured using AVGPR. 
To mitigate random noise in the results, for each dropout rate and case study ten equivalent models were trained and the resulting AVGPR were averaged. 
The results are shown in \autoref{fig:dropout_rates}.
Both for the Mnist and Cifar10 case studies, on both nominal and OOD test data and on all tested quantifiers,  AVGPR appears largely unimpacted by the dropout rate, as long as the dropout rate is less than or equal to $p=0.7$. 
On Cifar10, the decline is more continuous, starting already around $p=0.4$.
This might be caused by differing architectures, which clearly also have an impact on $p$'s influence. In fact,
the Cifar10 model has three dropout layers after convolutional layers, but none before the final output layer, while the
Mnist and Traffic models have dropout layers in both the convolutional and the dense parts of the model.

\autoref{fig:dropout_rates} shows that the dropout rate is not an overly critical hyperparameter to optimize uncertainty quantification, when chosen within reasonable bounds. 
Clearly, it shows that one should not artificially increase the dropout rate to ensure sufficient randomness for UQ. 
Indeed, while increasing the dropout rate might seem intuitively useful for the quantification of uncertainty under high variability, our results show that it can actually harm UQ performance if brought to an excessive level.

\vspace{0.5cm}
\begin{tcolorbox}
\textbf{Summary (RQ5)}:
\textit{There is no need to artificially increase the dropout rate to create randomness for UQ in MC-Dropout.
Within reasonable bounds, the UQ performance is only minimally sensitive to the choice of the dropout rate.
While the optimal rate might be model and dataset specific, 
any low dropout rate (in the range 0.2 - 0.3) performed generally well across models and datasets. } 
\end{tcolorbox}

\newcounter{guidelineCounter}
\stepcounter{guidelineCounter} %
\newcommand\guideline[1]{\colorbox{lightgray}{Guideline \theguidelineCounter \stepcounter{guidelineCounter}} #1}

\newpage
\subsection{Lessons Learned and Actionable Guidelines}
Based on our answers to the RQs, we distilled the following primary lessons learned and practical guidelines,
possibly useful for a practical usage of 
uncertainty-quantifiers when monitoring and supervising a DLS:

\begin{description}
    \item [The choice of which supervisor to adopt is not critical:]
    While the selection of the ideal supervisor is problem dependent, 
    all the supervisors we tested showed a good capability to increase the accuracy of the  DNN when supervised.
    Thus including any uncertainty monitoring supervisor in a DLS increases its fail-safety.
    
    \guideline{Use a supervisor, as it is always effective.}
    
    \item [Ensembles are powerful:]
    Not only did Ensembles show the best average rank on the $S_1$ score (ignoring the pre-trained Imagenet studies), in many cases even the unsupervised accuracy increased.
    Furthermore, the relatively low number of 20 atomic models was sufficient to achieve good results in our study.
    Thus, provided sufficient system resources, we suggest software architects to use Ensembles instead of Point Predictors. On the other hand, the latter may represent a good compromise solution when computational resources are severely constrained.
    
    \guideline{If enough computational resources are available, use Deep Ensembles.}
    
    \guideline{In more constrained environments, use MC-Dropout (fast training, slow predictions) or Proint Predictor based UQ (fast for training and prediction, but not theoretically well founded).}
    
    \item [Number of samples affects choice of quantifier:]
    For Ensembles, VR was the best quantifier on average, but it requires a large number of samples to allow for precise threshold selection. 
    Thus, if computational resources allow the training of a high number of atomic models, our experiments suggest to use VR as quantifier. Otherwise, MS showed good performance, despite its simplicity. 
    For MC-Dropout, where all quantifiers performed comparably well on average, one may prefer using PE, MI or AS that require fewer samples.
    
    \guideline{If you can afford to collect a (high) number of samples, choose the VR quantifier. Otherwise, try PE, MI and MS on 20 samples.}
    
    \guideline{With MC-Dropout, there appears not to be a clear advantage from using VR, which suggests the use of the non-discrete PE, MI or MS.}
    
    \guideline{Leverage \tool, which allows to use and compare multiple quantifiers at once at only marginal additional cost, to identify quantifiers which work well in your setting during alpha and beta testing.}
    
    \item [Developing PW-BNN is hard:] 
    Not only did the Flipout-Based PW-BNN we tested had a lower prediction performance then the corresponding PPNN with equivalent architecture,
    but we also found models to be highly sensitive to the choice of hyperparameters.
    At the same time, only PW-BNN allow to choose and interpret model properties that are specifically interesting for uncertainty quantification, such as the layers prior and posterior distribution.
    Thus, using PW-BNN in a safety-critical application requires even more extensive and costly hyperparameter tuning than the other approaches we tested.
    Based on our experiments, we recommend the use of PW-BNN only to developers with a well-founded understanding of uncertainty and sufficient time and resources to design, investigate and fine-tune their models extensively.
    
    \guideline{Only use Flipout-BNN if you know exactly what you are doing and need to take advantage of the advanced options provided by PW-BNN (such as extracting a posterior distribution).}
    
    \guideline{Never blindly copy-paste a PPNN architecture and replace its layers with Flipout layers.}
    
    \guideline{If you use Flipout-BNN, start by making only the last layer of your model probabilistic or
    be prepared for an exhaustive hyperparameter search. }
    
    \item [MC-Dropout is only mildly sensitive to dropout rate:] 
    Within reasonable bounds, the dropout rate used for training only mildly influenced the performance of MC-Dropout and there was no dropout rate that performed best over all models.
    Hence, we do not suggest a specific dropout rate to be used by default and do not deem the dropout rate to be a critical hyperparameter in MC-Dropout in most cases, if chosen within a reasonable range, such as 0.2-0.3.
    
    \guideline{Do not add extra randomness (i.e., excessively high dropout rate) to your model for MC-Dropout. It is not needed.}
    
    \item [In-production supervisor assessment is needed:]
    Since there is no uncertainty quantifier that performs best in all cases,
    we want to emphasize the importance of in-production assessment of the supervisors performance on the actual system to be supervised, by comparing different supervisors for optimal selection. The metrics that we propose in Section~\ref{sec:assessment} are specifically designed for such assessment.
    
    \guideline{Monitor, evaluate and adapt your supervisor iteratively.}
\end{description}

Here, we want to put special emphasis on the last guideline. 
While our recommendations and findings are based on a thorough and large-scale empirical evaluation, 
it lies in the nature of uncertainty  being constantly faced with previously unseen and untested conditions, 
to which no study can ever fully generalize.
Thus, system developers must  make sure they find the best supervisor choice and configuration for their systems, by iteratively experimenting and improving.
For this purpose, \tool may be helpful, as by modifying only a few lines of code, one can change the choice of uncertainty-aware DNN and critical hyperparameters, such as the used quantifiers and number of samples, and even use multiple quantifiers simultaneously.

\subsection{Threats to Validity}

\textbf{External Validity}: While we considered only four subjects, we diversified them as much as possible. In particular, besides the benchmark subjects often used in DNN testing (Mnist, Cifar10 and Imagenet), we included an additional subject, Traffic,  which consists of unbalanced data, partially of low quality. It implements a functionality (traffic sign recognition) commonly integrated in autonomous vehicles. 

\textbf{Internal Validity}: 
The selection of hyperparameters for DNN training
might be critical and the selected values may not be representative of other contexts.
To address this threat, we refrained from selecting any  hyperparameter for the case study models ourselves, wherever possible, and instead relied on  architectures available from the literature.
For what concerns the internal hyperparameters of the supervisors, we evaluated the sensitivity of the results to their choice in a dedicated research question (RQ4).

Another threat to the internal validity of our study is that the OOD inputs used in our experiments might not be representative of the uncertainties that may be observed in practice. 
Indeed, this is unavoidable and intrinsic to the problem of DNN supervision, as unexpected conditions occurring exclusively in practice cannot be by definition simulated ahead of time.

\section{Related Work}
\label{sec:related}

\textbf{Empirical Studies of Uncertainty-Aware Deep Neural Networks:}
Oliveira \etal \cite{Oliveira2016} compared, amongst others, MC-Dropout based uncertainty and a variational approximation of BNN. 
In their experiments, the performance of the two were comparable, but MC-Dropout was much faster. 
They did not consider Ensemble models.
Similarly, but more extensively, Ovadia \etal \cite{Ovadia2019} compared various uncertainty aware DNNs against each other, including MC-Dropout,
Deep Ensembles and a variational approximation of BNN. Consistently with our results,  Ensembles performed the best in their experiments.
As opposed to our work, both of these studies consider fewer  subjects, do not investigate the impact of different quantifiers,
and do not have the constraint of transparently introducing uncertainty estimators into DNNs without altering their inner architecture, as possible instead with variational BNN approximations.
Zhang \etal \cite{Zhang2020} compare MC-Dropout against PCS to detect misclassifications caused by adversarial examples,
i.e., examples deliberately modified to trick the DNN into making prediction errors. 
Their results show, similarly to ours, that there is no strict dominance between these two approaches to network supervision. Ours is the first large scale study where uncertainty estimators are injected transparently into existing DNNs (thanks to \tool). We are also the first to introduce practical metrics for in-production assessment of supervisors and to distill a list of lessons learned that can be used as guidelines for practical usage of supervisors.

\textbf{Other types of DNN supervisors:}
While our focus is on uncertainty measures based on the variability of the output for a given input,
there have been various proposals of supervisors that are not directly based on the output distribution of the supervised network.
Berend \etal\cite{Berend2020} compared various such techniques, typically based on neuron activations, which re\-co\-gnize activation patterns that were not sufficiently represented in the training data. While such an approach may be powerful against epistemic uncertainty, it can not help against aleatoric uncertainty.
Stocco \etal\cite{Stocco2020} and Henriksson \etal\cite{Henriksson2019} proposed the use of autoencoders, i.e., anomaly detectors trained on the DLS training set as DNN supervisors. This approach does not consider the inner state of the DNN or its predictions (i.e., it is black-box). On the contrary, the supervisors supported by \tool take advantage of the predictions of the DNN being supervised.

\section{Conclusion}
\label{sec:conclusion}

Uncertainty quantification is fundamental to ensure fail-safe execution of DLS, as uncertainty quantifiers support the implementation of supervisors that can disengage the DNN and heal the system when predictions are unreliable. The goal of our work was to simplify the adoption of supervisors in DLS by providing developers with information about the comparative differences among existing uncertainty estimators proposed in the literature, by investigating the sensitivity of such approaches to hyperparameters, and by distilling our lessons learned into actionable guidelines. Among such guidelines, key ones are the noncritical choice of which supervisor to use, the advantages offered by Ensemble supervisors, when enough computational resources allow their adoption, and in general the importance of in-field assessment and improvement of the adopted supervisors. The latter activities  are highly facilitated by our tool \tool and by the assessment framework described in this paper.

Among the future works enabled by the research presented in this paper, the hybridization of different supervisors is definitely very appealing and promising, with the ultimate goal of further simplifying the adoption and optimal configuration of a supervisor by developers.

\section*{Acknowledgments}

\subsection*{Financial disclosure}

This work was partially supported by the H2020 project PRECRIME, funded under the ERC Advanced Grant 2017 Program (ERC Grant Agreement n. 787703). 

\subsection*{Conflict of interest}

The authors declare no potential conflict of interests.

\section*{Supporting information}
The replication package for our experiments is released under a permissive MIT license and can be found on \[\text{\href{https://github.com/testingautomated-usi/repli-stvr2021-uncertainty}{https://github.com/testingautomated-usi/repli-stvr2022-uncertainty}}.\]
It is also publicly archived on zenodo, where it can be found on \[\text{\href{https://zenodo.org/badge/latestdoi/413278605}{https://zenodo.org/badge/latestdoi/413278605}}.\]

\bibliography{main}%

\newpage
\section*{Author Biography}

\begin{biography}{\includegraphics[width=66pt,height=86pt]{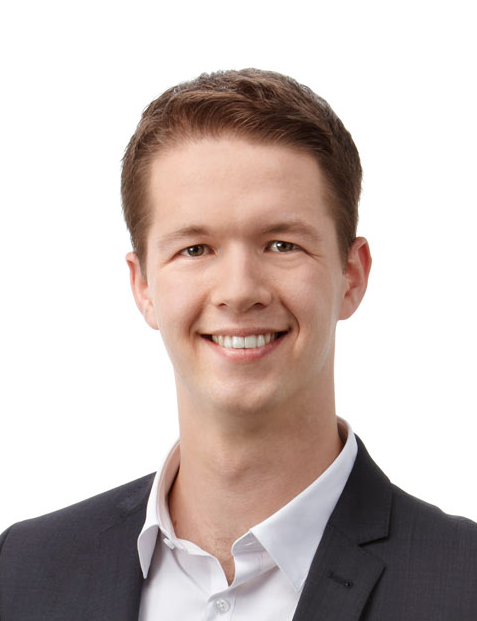}}{\textbf{Michael Weiss.} Michael Weiss is a Phd candidate at the Software Institute, Università della Svizzera italiana (USI). He co-authored various papers at top tier artificial intelligence and software engineering venues. His research interests include software engineering and software robustness in systems that rely on artificial intelligence.
He focuses on designing fail-safe approaches for machine learning based systems, as well as on the development of tools to facilitate the use of such approaches for developers.
He is the principal author of \tool, an award-winning python library allowing fast and near-transparent uncertainty quantification of deep neural networks.}
\end{biography}

\begin{biography}{\includegraphics[width=66pt,height=86pt]{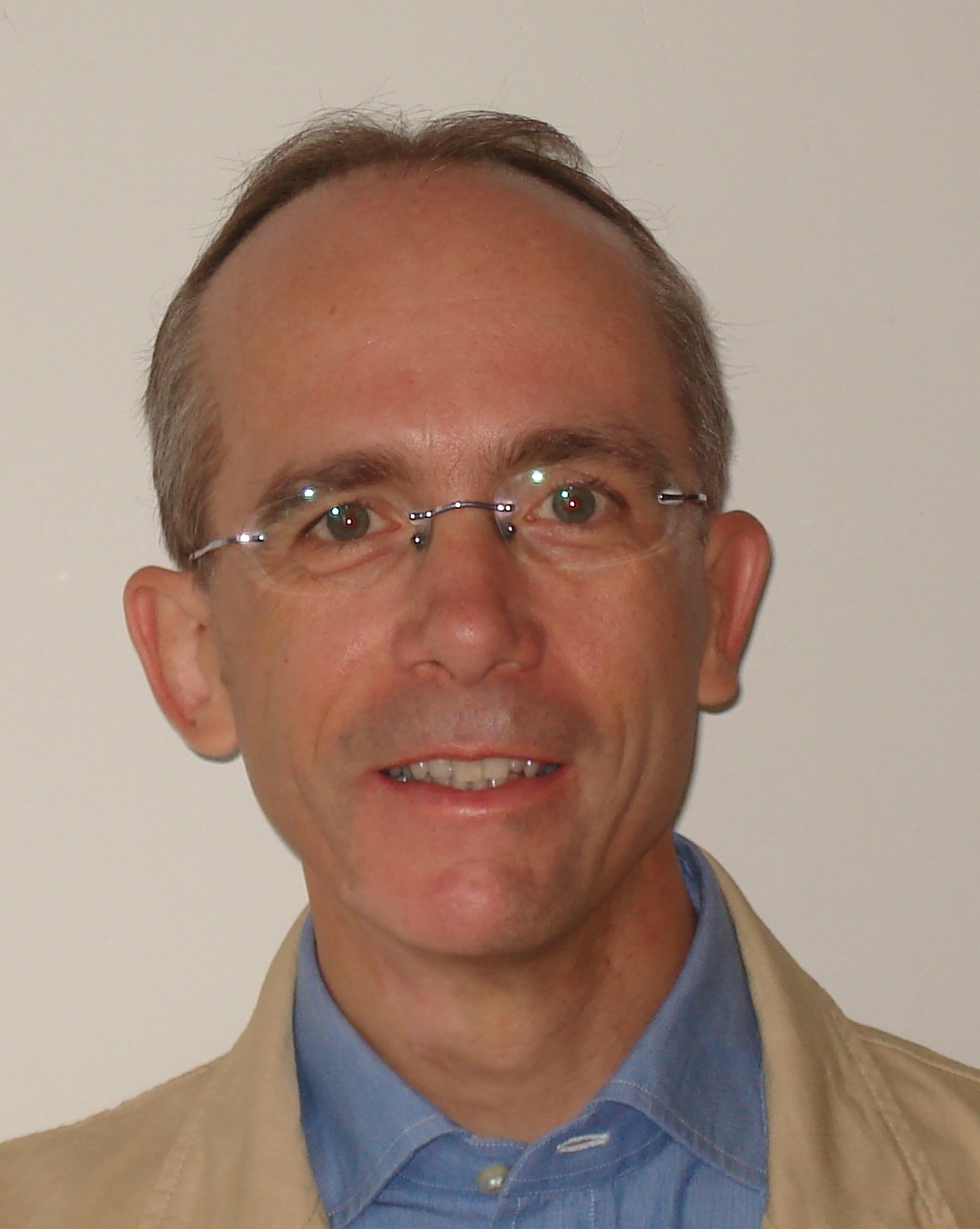}}{\textbf{Paolo Tonella.} Paolo Tonella is Full Professor at the Faculty of Informatics and at the Software Institute of Universita' della Svizzera Italiana (USI) in Lugano, Switzerland. He is also Honorary Professor at University College London, UK. Until mid 2018 he has been Head of Software Engineering at Fondazione Bruno Kessler, Trento, Italy. Paolo Tonella is the recipient of an ERC Advanced grant as Principal Investigator of the project PRECRIME. In 2011 he was awarded the ICSE 2001 MIP (Most Influential Paper) award, for his paper: "Analysis and Testing of Web Applications". He is the author of "Reverse Engineering of Object Oriented Code", Springer, 2005, and of "Evolutionary Testing of Classes", ISSTA 2004. Paolo Tonella was Program Chair of ICSM 2011 and ICPC 2007; General Chair of ISSTA 2010 and ICSM 2012. He is/was associate editor of TOSEM/TSE and he is in the editorial board of EMSE and JSEP. His current research interests include deep learning testing, web testing, search based test case generation and the test oracle problem.}
\end{biography}

\appendix

\section{Training the Flipout Model for the Traffic Case Study}
\label{sec:appendix_traffic_bnn}

Training the Flipout PW-BNN for \autoref{tab:big_table} required a number of attempts until a valid and useful model could be collected:
On the first tries, the model (with the same code that was used to later train one model successfully) would  perform far below expectation, with an accuracy of only $~6\%$.
In this Section, we discuss where this suboptimality came from and how it relates to the chosen hyperparameters. 
To investigate this issue, we trained 100 additional Flipout based models, 10 per setting:
\begin{itemize}
    \item \textbf{Setting 0:} The equivalent hyperparameters as used in our traffic PPNN, with all convolutional and dense layers replaced by Flipout layers and all dropout layers removed. This is also the setting used in \autoref{tab:big_table}. 
    \item \textbf{Setting 1-4:} Same as Setting 0, with one parameter of the optimizer changed, or a change of optimizer from SGD to RMSProp (using the RPMSProp default configuration).
    \item \textbf{Setting 5-9:} Same as Setting 0, but not all of the six convolutional and two dense layers were implemented as flipout layers. 
\end{itemize} 

All models were trained for 100 epochs and evaluated on the nominal validation set, where predictions were made single-shot, i.e., only a single sample was collected per prediction, as is common to monitor progress during model training. 
The resulting final accuracies are reported in \autoref{tab:flipout_traffic_search}. 
Values where training clearly failed are highlighted yellow. 
This table and the corresponding training logs allow interesting observations about the difficulties which occurred during training:

\begin{table}[b]
\centering

\scriptsize
\begin{tabular}{@{}cccccccccccccc@{}}
\toprule
Setting & \begin{tabular}[c]{@{}c@{}}\#Flipout-L \\ (Conv. / Dense.)\end{tabular} & \begin{tabular}[c]{@{}c@{}}Optimizer\\ (Alg / Rate / Mom.)\end{tabular} & \multicolumn{10}{c}{\begin{tabular}[c]{@{}c@{}}Accuracy\\ (on 10 runs)\end{tabular}} \\ 
\midrule
        &                                                                              &                                                                       & 0        & 1        & 2        & 3        & 4        & 5        & 6        & 7   & 8 & 9       \\
0&6  / 2 &SGD* / 0.01* / - *  & 0.91& 0.93& 0.93& 0.94& \hl{0.06}& 0.92& 0.94& 0.93& \hl{0.7}& \hl{0.06}\\
1&6  / 2 &SGD* / 0.05 / - *  & \hl{NaN}& \hl{NaN}& \hl{NaN}& \hl{NaN}& \hl{NaN}& \hl{NaN}& \hl{NaN}& \hl{NaN}& \hl{NaN}& \hl{NaN}\\
2&6  / 2 &SGD* / 0.1 / - *  & \hl{NaN}& \hl{NaN}& \hl{NaN}& \hl{NaN}& \hl{NaN}& \hl{NaN}& \hl{NaN}& \hl{NaN}& \hl{NaN}& \hl{NaN}\\
3&6  / 2 &SGD* / 0.01* / 0.1  & 0.93& 0.93& 0.93& 0.94& 0.93& 0.93& 0.92& 0.93& \hl{0.07}& 0.91\\
4&6  / 2 &RMSProp / 0.001** / - **  & \hl{0.06}& \hl{0.05}& \hl{0.06}& \hl{0.06}& \hl{0.06}& \hl{0.06}& \hl{0.06}& \hl{0.06}& \hl{0.06}& \hl{0.06}\\
5&4  / 2 &SGD* / 0.01* / - *  & 0.94& \hl{0.06}& \hl{0.06}& 0.92& \hl{0.06}& \hl{0.06}& 0.91& \hl{0.06}& 0.94& \hl{0.05}\\
6&2  / 2 &SGD* / 0.01* / - *  & 0.96& 0.96& 0.96& 0.96& 0.96& 0.96& 0.96& 0.96& 0.96& 0.96\\
7&6  / 1 &SGD* / 0.01* / - *  & 0.96& 0.97& 0.97& 0.97& 0.97& 0.97& 0.97& 0.97& 0.96& 0.97\\
8&4  / 1 &SGD* / 0.01* / - *  & 0.97& 0.97& 0.97& 0.97& 0.97& 0.97& 0.97& 0.97& 0.97& 0.97\\
9&2  / 1 &SGD* / 0.01* / - *  & 0.98& 0.98& 0.98& 0.98& 0.98& 0.98& 0.98& 0.98& 0.98& 0.98\\
\bottomrule
&&&&&&&&\multicolumn{2}{c}{\footnotesize{* default}}&\multicolumn{3}{r}{\footnotesize{** default for RMSProp}}
\end{tabular}
\caption{Flipout model accuracy after training, using different hyperparameters.}
\label{tab:flipout_traffic_search}
\end{table}

First, we observe that with settings 0, 4, and 5, a large number of training runs failed to achieve a higher accuracy than 6\%. 
Inspecting the corresponding training logs, we observe that while the loss is continuously decreasing, it does so only very slowly, such that after 100 epochs the model is nowhere near optimal. 
An example of such a run is provided in \autoref{fig:flipout_tensorboard} (a and d), showing validation loss and accuracy during training for Setting 0 in run 4, where the loss only decreased very mildly from ~100 to ~99.5, where good models typically achieved a loss around 95. 
Actually, we sometimes even observe in successful training runs such as Setting 5 run 6, shown in \autoref{fig:flipout_tensorboard} (c and f), that for the first 41 epochs, the model did not make any observable progress, before finally descending heavily in loss. 
While this seems like a minor issue, it is practically highly relevant: 
In much larger, industry-level models, training can take days, and not seeing any notable improvements during these days makes debugging much harder - especially when, as we see for the remaining entries of Setting 5, in many cases the model actually does not suddenly improve.
This slow learning could potentially be caused by the initial \emph{burn-in time} that is often needed in MCMC appraoches~\cite{Jospin2020},
but has also been shown for other types of PW-BNN, such as for Stochastic Variational Inference~\cite{Rossi2019}.

One may try to tackle the problem of slow progress by choosing higher learning rates (Settings 1 and 2) or by using momentum in the optimizer (Setting 3).
However, higher learning rates quickly led to \emph{Not-a-Number (NaN)} errors, likely caused by exploding gradients, eventually crashing the training process,
when, in our experiments, the small momentum used in Setting 3 actually appeared to be a sweet-spot setting, where 9 out of 10 runs completed successfully.
The fact that the range between too slow learning (no convergence in reasonable time) and too fast learning (exploding gradients) is very tight, is best highlighted by Setting 6:
While we above considered the learning rate in run 6 as too small, the same setting led to an exploding gradient in run 4 (shown in \autoref{fig:flipout_tensorboard} b and e), leading to a peak loss of more than 1200. 
As a last interesting observation, we see that in Settings 6 to 9, where not all convolutional and dense layers were implemented as Flipout layers, training was clearly much more successful, indicating that the difficulties are indeed related to the Flipout layers.

From these experiments, we learn that clearly one should not just take a PPNN and replace its layers with Flipout layers to allow for uncertainty quantification. 
When using Flipout layers, the choice of hyperparameters appears to be even more critical than for straightforward PPNN, making their training a much more complex task.
Here, it is worth mentioning that despite the discouraging results we found in our experiments, we do not consider Flipout layers, or PW-BNN in general as useless or bad, as they also come with a very distinct advantage that goes beyond the scope of this paper:
PW-BNNs allow specifying a prior distribution, and to observe a posterior distribution which allows the skilled UQ researchers to encode and extract much more specific assumptions and information than with other approaches such as PPNN or Deep Ensembles. 
Carefully designed and finetuned, PW-BNN may then also become competitive with the simpler approaches in terms of prediction accuracy and UQ capability.
An intermediate, easier-to-design solution might be to use Flipout only for some layers, for which our experiments suggest a higher training success rate. 
This idea is also motivated by existing research: Riquelme \etal ~propose to use a bayesian linear regression only in a model's last layer, which represents "a robust and easy-to-tune approach"~\cite{Riquelme2018}. 
Our empirical evaluation clearly supports such a claim for Flipout layers.

\def \bnntraffsubfigurewidth {0.33\linewidth}
\def \bnntraffimgwidth {.9\linewidth}

\begin{figure}[t]
\centering
\begin{subfigure}{\bnntraffsubfigurewidth}
  \centering
  \includegraphics[width=\bnntraffimgwidth]{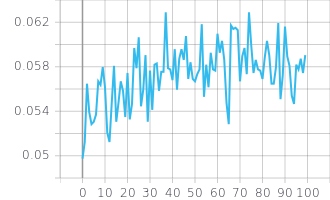}
  \caption{\textbf{slow progress} (accuracy)}
\end{subfigure}%
\begin{subfigure}{\bnntraffsubfigurewidth}
  \centering
  \includegraphics[width=\bnntraffimgwidth]{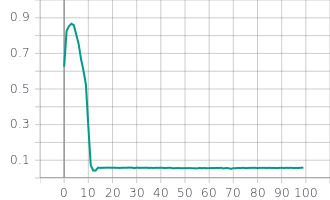}
  \caption{\textbf{exploding gradient?} (accuracy)}
\end{subfigure}
\begin{subfigure}{\bnntraffsubfigurewidth}
  \centering
  \includegraphics[width=\bnntraffimgwidth]{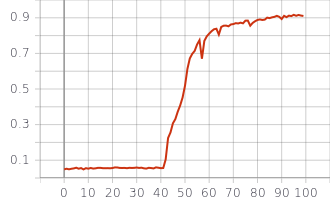}
  \caption{\textbf{late improvements} (accuracy)}
\end{subfigure}

\begin{subfigure}{\bnntraffsubfigurewidth}
  \centering
  \includegraphics[width=\bnntraffimgwidth]{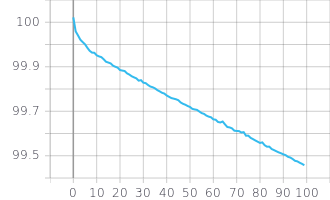}
  \caption{\textbf{slow progress} (loss)}
\end{subfigure}%
\begin{subfigure}{\bnntraffsubfigurewidth}
  \centering
  \includegraphics[width=\bnntraffimgwidth]{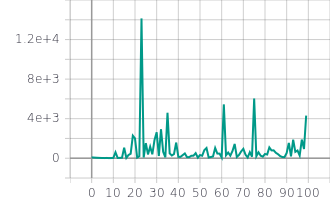}
  \caption{\textbf{exploding gradient?} (loss)}
\end{subfigure}
\begin{subfigure}{\bnntraffsubfigurewidth}
  \centering
  \includegraphics[width=\bnntraffimgwidth]{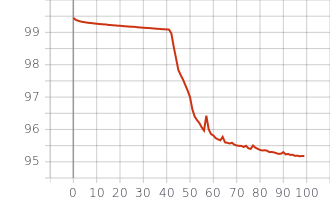}
  \caption{\textbf{late improvements} (loss)}
\end{subfigure}

\caption{Examples of problems faced when training the flipout based PW-BNN for the traffic case study}
\label{fig:flipout_tensorboard}
\end{figure}

\end{document}